\documentclass[preprint2]{aastex62}
\usepackage{float}
\usepackage{graphicx}

\newcommand{\og}{2014~OG$_{392}$}
\newcommand{\ogColor}{$1.64\pm0.4$} 

\begin{document} 

\title{Cometary Activity Discovered on a Distant Centaur: A Non-Aqueous Sublimation Mechanism}

\shorttitle{Centaur Activity Discovery and Non-Aqueous Sublimation}
\shortauthors{Chandler et al.}

\correspondingauthor{Colin Orion Chandler}
\email{orion@nau.edu}

\author[0000-0001-7335-1715]{Colin Orion Chandler}
\affiliation{Department of Physics \& Astronomy, Northern Arizona University, PO Box 6010, Flagstaff, AZ 86011, USA}

\author[0000-0001-8531-038X]{Jay K. Kueny}
\affiliation{Department of Physics \& Astronomy, Northern Arizona University, PO   
Box 6010, Flagstaff, AZ 86011, USA}
\affiliation{Lowell Observatory, 1400 W Mars Hill Rd, Flagstaff, AZ 86001, USA}

\author[0000-0001-9859-0894]{Chadwick A. Trujillo}
\affiliation{Department of Physics \& Astronomy, Northern Arizona University, PO Box 6010, Flagstaff, AZ 86011, USA}

\author[0000-0003-4580-3790]{David E. Trilling}
\affiliation{Department of Physics \& Astronomy, Northern Arizona University, PO Box 6010, Flagstaff, AZ 86011, USA}

\author[0000-0001-5750-4953]{William J. Oldroyd}
\affiliation{Department of Physics \& Astronomy, Northern Arizona University, PO Box 6010, Flagstaff, AZ 86011, USA}

\begin{abstract}
Centaurs are minor planets thought to have originated in the outer Solar System region known as the Kuiper Belt. Active Centaurs enigmatically display comet-like features (e.g., tails, comae) even though they orbit in the gas giant region where it is too cold for water to readily sublimate. Only 18 active Centaurs have been identified since 1927 and, consequently, the underlying activity mechanism(s) have remained largely unknown up to this point. Here we report the discovery of activity emanating from Centaur \og{}, based on archival images we uncovered plus our own new observational evidence acquired with the Dark Energy Camera (Cerro Tololo Inter-American Observatory Blanco 4~m telescope), the Inamori-Magellan Areal Camera \& Spectrograph (Las Campanas Observatory 6.5~m Walter Baade Telescope) and the Large Monolithic Imager (Lowell Observatory 4.3~m Discovery Channel Telescope). We detect a coma as far as 400,000 km from \og{}, and our novel analysis of sublimation processes and dynamical lifetime suggest carbon dioxide and/or ammonia are the most likely candidates for causing activity on this and other active Centaurs. We find \og{} is optically red, but CO$_2$ and NH$_3$ are spectrally neutral in this wavelength regime so the reddening agent is as yet unidentified.
\end{abstract}


\section{Introduction}
\label{sec:introduction}

Prior to the mid-twentieth century, comets were thought to be the only astronomical objects with tails or comae. Unsurprisingly, then, the first two active Centaur discoveries -- 29P/Schwassman-Wachmann~1 \citep{SW1927SW1} and 39P/Oterma \citep{Oterma1942CometOterma} -- were initially classified as comets.

\begin{figure}[H]
    \centering
	\includegraphics[width=0.8\columnwidth]{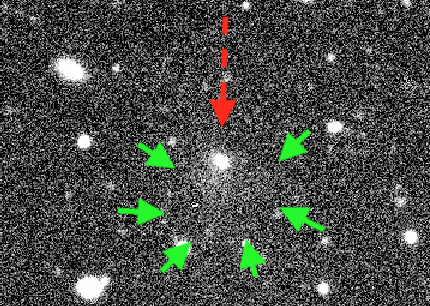}
	\caption{\og{} (dashed arrow) displays a coma (short arrows) during our August 30 2019 observations. Stack of $4\times250$~s DECam exposures.} 
	\label{fig:stacked}
\end{figure}

\noindent In 1949 the discovery of the first Active Asteroid, (4015)~Wilson-Harrington (also designated 107P), blurred the dividing line between asteroid and comet \citep{r02043}. In 1977 (2060)~Chiron was discovered \citep{1977IAUC.3129....1K}, the first member of the population now known as Centaurs. (2060)~Chiron was later found to be active, making it the first object to be identified as a Centaur prior to activity discovery \citep{Meech1990ChironAtmosphere}.

We adopt the Centaur classification system \citep{Jewitt2009TheActiveCentaurs} which defines Centaurs as objects (1) with perihelia and semi-major axes between the orbits of Jupiter ($\sim$5 au) and Neptune ($\sim$30~au) and (2) not in 1:1 mean-motion-resonance with a giant planet (as is the case for the Trojans). We distinguish between Centaurs and Jupiter Family Comets (following \citealt{1994Icar..108...18L}) via the Tisserand parameter with respect to Jupiter, given by

\begin{equation}
	T_\mathrm{J} = \frac{a_\mathrm{J}}{a} + 2\sqrt{\left(1-e^2\right)\frac{a}{a_\mathrm{J}}}\cos(i),
	\label{eq:TJ}
\end{equation}

\noindent with eccentricity $e$, inclination $i$, and the semi-major axes of the body and Jupiter $a$ and $a_\mathrm{J}$, respectively. Centaurs have $T_\mathrm{J}>=3$ whereas Jupiter Family Comets range between $2<T_\mathrm{J}<3$.

\begin{table*}
    \centering
    \small
	\caption{Active Centaurs}
	\label{tab:activecentaurs}
	\hspace{45mm}\textit{-- Orbital Elements --}\hspace{16mm}\textit{-- Activity Discovery --}\\
	\begin{tabular}{l|rrrr|crccr}
		Object Name or Designation & $P$  & $a$  & $q$  & $Q$  & $r$  & $\%_{T\rightarrow q}$	& $M_V$	& Date & Ref.\\
									& [yr] & [au] & [au] & [au] & [au] & & & [UT] & \\
		\hline
		Chiron~(95P)				& 50.5	& \ 6.0 &  8.5 & 18.9	& 11.8  & 68 	& 17.0 & 1989-04-10 & 1\\
		Echeclus~(174P)				& 35.3	&  10.8 &  5.9 & 15.6	& 13.1	& 25	& 21.1	& 2005-12-04 & 2\\
		29P/Schwassmann-Wachmann 1	& 14.8	& \ 6.0 &  5.5$^\dagger$ & \ 6.6    & \ 6.0	& 53 & 15.3	& 1927-11-15 & 3\\
		39P/Oterma					& 19.5	& \ 7.2 &  3.4$^\dagger$ & \ 9.0	& \ 3.5	& 99 & 15.1 & 1942-02-12 & 4\\
		165P/LINEAR					& 76.4	&  18.0 &  6.8 & 29.3	& \ 6.9	& 100	& 19.4	& 2000-01-09 & 5\\
		166P/NEAT					& 51.9	&  13.9 &  8.6 & 19.2	& \ 8.6	& 100	& 19.6	& 2001-10-15 & 6\\
		167P/CINEOS					& 64.8	&  16.1 & 11.8 & 20.5	&  12.2 & 96	& 20.7	& 2004-06-07 & 7\\
		P/2005~S${2}$ (Skiff)		& \ 22.5 	& \ 8.0 & \ 6.4 & \ 9.5	& \ 6.5	& 98	& 19.7	& 2005-09-16 & 8\\
		P/2005~T$_{3}$ (Read)		& \	20.6	& \ 7.5 & \ 6.2 & \ 8.8	& \ 6.2	& 100	& 20.7	& 2005-10-07 & 9\\
		P/2011~C${2}$ (Gibbs)		& \ 20.0	& \ 7.4 & \ 5.4 & \ 9.3	&  \ 5.5	& 97	&   20.3	&2011-02-12 & 10\\
		C/2011~P${2}$ (PanSTARRS)	& \ 30.6	& \ 9.8 & \ 6.2 & 13.4	&  \ 6.3	&  98	&   20.3	& 2011-08-03 & 11\\
		P/2011~S${1}$ (Gibbs)		& \ 25.4	& \ 8.6 & \ 6.9 & 10.4	&  \ 7.5	&   82	&   21.0	&2011-09-18 & 12\\
		C/2013~C${2}$ (Tenagra)	& \ 64.4	&  16.1 & \ 9.1 & 23.0	& \ 9.8	&   96	&   19.1	&2013-02-14 & 13\\
		C/2013~P${4}$ (PanSTARRS)	& \ 56.8	&  14.8 & \ 6.0 & 23.6	&  \ 6.3	&   98	&   19.5	&2013-08-15 & 14\\
		P/2015~M${2}$ (PanSTARRS)	& 19.3     & \ 7.2  & \ 5.9& \ 8.5	& \ 5.9    &   100 &   19.5	& 2015-06-28 & 15\\
		C/2015~T${5}$ (Sheppard-Tholen)& 147.9	&  28.0 & \ 9.3 &   46.6	&  \ 9.4	&   100	&   22.3	&2015-10-13 & 16\\
		C/2016~Q${4}$ (Kowalski)	& \ 69.0	&  16.8 & \ 7.1 & 26.5	&  \ 7.5	&   98	& 20.1	& 2016-08-30 & 17\\
		2003~QD$_{112}$				& \ 82.8	&  19.0 & \ 7.9 & 30.1	&   12.7 	&   57	&   21.7     & 2004-10-10 & 18\\
		\og{}				& \ 42.5	&  12.2 & \ 10.0 & 14.4	& 10.6	&   86 &21.1  & 2017-07-18 & 19\\
	\end{tabular}
	
	\raggedright
	$P$: orbital period; $a$: semi major axis; $q$: perihelion distance; $Q$: aphelion distance; $r$: heliocentric distance; $\%_{T\rightarrow q}$: fractional perihelion-aphelion distance (Equation \ref{eq:percentperi}); $M_V$: apparent $V$-band magnitude. $Q$ computed via $Q=a(1+e)$ when otherwise unavailable. Asteroid parameters provided by the Minor Planet Center. Heliocentric distance and apparent magnitude courtesy of JPL Horizons \citep{Giorgini1996Horizons}.

    $^\dagger$ original value(s) from activity discovery epoch adopted where available; otherwise values adopted from more recent epoch(s). Ref. points to a source which discusses activity of the object.
    
    References: 1:\cite{Meech1990ChironAtmosphere}, 2:\cite{2006IAUC.8656....2C}; 3:\cite{SW1927SW1}, 4:\cite{Oterma1942CometOterma}, 5:\cite{2005IAUC.8552....2G}, 6:\cite{2001IAUC.7738....1P}, 7:\cite{Romanishin2004act167P}, 8:\cite{Gajdos2005discP2005S2Skiff}, 9:\cite{Read2005P2005T3IAUC}, 10:\cite{Gibbs2011P2011C2IAUC}, 11:\cite{Wainscoat2011P2011P2IAUC}, 12:\cite{Gibbs2011C2011S1IAUC}, 13:\cite{Holvorcem2013C2013C2CBET}, 14:\cite{Wainscoat2013C2013P4CBET}, 15:\cite{Bacci2015P2015M2MPEC}, 16:\cite{Tholen2015P2015T5CBET}, 17:\cite{Kowalski2016C2016Q4activity}, 18:\cite{Jewitt2009TheActiveCentaurs}, 19:this work
\end{table*}

Centaurs are thought to have migrated inwards from the Kuiper Belt (see review by \citealt{2008tnoc.book...79M}), a region that spans 30~au (Neptune's orbital distance) to 50~au. Neptune Trojans may also serve as a Centaur reservoir \citep{2010MNRAS.402...13H}. Centaurs all orbit exterior to the 3~au water ice line so they cannot readily undergo sublimation. Surprisingly, though, 18 Centaurs ($\sim$ 4\% of known Centaurs) have been found to display prominent comet-like features such as comae (e.g., Fig. \ref{fig:stacked}) or tails; these are the active Centaurs. Table \ref{tab:activecentaurs} lists the known active Centaurs along with key physical parameters and discovery circumstances.

Our understanding of active Centaurs has been limited because of their faint apparent magnitudes (the mean apparent magnitude $m_V$ at discovery is $\sim$20; Table \ref{tab:activecentaurs}), since it is necessary to probe several magnitudes fainter in order to reliably detect activity via telescopic imaging. Spectroscopy has been used with some success to identify cometary activity originating from asteroids \citep{2018Icar..304...83B} but this method requires even brighter targets than detection by imaging. Discovering activity on Centaurs is observationally challenging because they are faint, telescope time-intensive, and because they are rare. Active centaurs are discovered, on average, within $\sim$10\% of their perihelion distance (Table \ref{tab:activecentaurs}) where they are significantly brighter and, importantly, warmer.

Another significant obstacle to understanding active Centaurs stems from the extreme cold found at their orbital distances. Water and methanol ices have been detected on the surfaces of $\sim$10 Centaurs, but only one of these, (2060)~Chiron, has also been visibly active (see review, \citealt{Peixinho2020CentaursReview}). At surface temperatures less than 150~K and pressures below $\sim10^{-12}$ bar many thermodynamical properties (e.g., enthalpy of sublimation) of volatile ices are not well known from laboratory experiments \citep{WOS:000273099100041}. Moreover, ices may exist in two or more different structural forms; energy from the H$_2$O crystalline--amorphous state transition may even play a role in generating activity \citep{Jewitt2009TheActiveCentaurs}.


\section{Mining Archival Data}
\label{sec:archivaldata}

In order to overcome the observational challenges discussed in Section \ref{sec:introduction} we began by searching archival images captured with the 0.5~gigapixel Dark Energy Camera (DECam) on the Blanco 4~m telescope at the Cerro Tololo Inter-American Observatory in Chile. Archival data from this facility allow the detection of faint activity because of the relatively large aperture and because a large number of objects serendipitously imaged by the instrument can be searched.

We identified Centaurs in our own proprietary database cataloging the NSF’s National Optical-Infrared Astronomy Research Laboratory (NSF’s OIR Lab, formerly NOAO) public DECam archive following the methodology outlined in \cite{Chandler2018SAFARI}. Our general approach was to correlate image celestial coordinate and temporal data with object ephemeris services such as NASA JPL Horizons \citep{Giorgini1996Horizons} and IMCCE SkyBot (\citet{Berthier:2006tn}; see also the acknowledgements).

We (1) extracted event information from the entire DECam public archive database, (2) submitted objects to SkyBot or matched against ephemerides produced via the Minor Planet Center and/or Horizons, and then (3) carried out a database query to identify potential images containing Centaurs. 

After (4) downloading the data, we (5) checked each chip for the presence of the Centaur to ensure the object was visible and free of imaging complications (e.g., gaps between chips, scattered light from bright stars, cosmic rays). Finally, we (6) adhered to the routine outlined in \cite{Chandler2018SAFARI} where, following image file retrieval of \og{} from the archive, we extracted Flexible Image Transport System (FITS) and Portable Network Graphics (PNG) thumbnails (480$\times$480 pixel images). We subjected these thumbnails to image processing techniques in order to assist by-eye analysis.

While examining each Centaur PNG thumbnail image by eye we flagged any with apparent activity for later analysis. FITS thumbnail images corresponding to those flagged were subjected to additional image processing techniques in an effort to enhance image quality, especially comae contrast.

To ascertain potential heliocentric distance effects we made use of a simple metric \citep{Chandler2018SAFARI}, $\%_{T\rightarrow q}$, which describes how close to perihelion ($q$) an object's distance ($d$) is relative to its aphelion distance ($Q$):

\begin{equation}
		\%_{T\rightarrow q} = \left(\frac{Q - d}{Q-q}\right)\cdot 100\mathrm{\%}.
		\label{eq:percentperi}
\end{equation}

\begin{figure*}
	\includegraphics[width=1\linewidth]{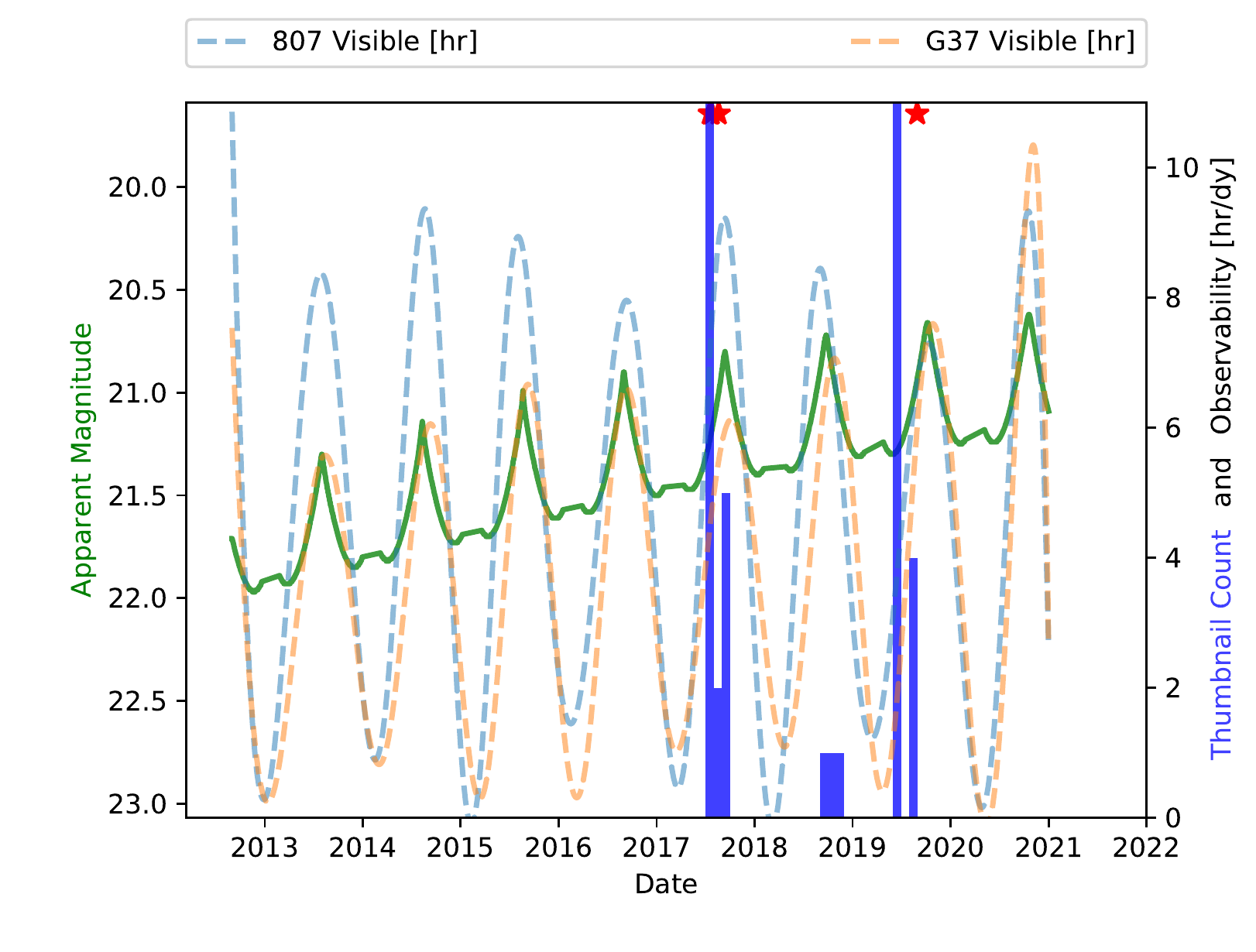}
	\caption{\og{} activity timeline beginning 2012 September (DECam first light) to present. Red stars show when we found visible activity. The orbital period is $\sim$42 years so neither perihelion (2021 December 3) nor aphelion are visible on this plot. The solid green line (left vertical axis) shows the geocentric apparent $V$-band magnitude of \og{}. Dashed lines (right vertical axis) indicate the number of nighttime hours with elevation $> 15^\circ$ for the southern hemisphere DECam (blue; site code: 807) and for the northern hemisphere Discovery Channel Telescope (orange; site code: G37).  The overlaid histogram (vertical blue bars and right axis) shows the number of thumbnail images captured during one calendar month. Note that in all instances when observability was high and many thumbnails were present, activity was observed.
    }
	\label{fig:ActivityTimeline}
\end{figure*}

From DECam archival data we extracted $\sim20$ thumbnail images of \og{}; Figure \ref{fig:ActivityTimeline} shows the number of thumbnails obtained along with the predicted apparent $V$-band magnitude and observability of \og{}. In images from July and August, 2017, we spotted what appeared to be activity emanating from \og{} (see gallery, Appendix \ref{sec:archivalimages}); at that time the object was 10.60~au from the Sun.

\section{Followup Observing}
\label{sec:observations}

To confirm the presence of activity we used the same DECam instrument and made additional observations on UT 30 August 2019. Fig.\ 1 shows \og{} with a telltale coma revealed by a combined 1000~s exposure. Appendix \ref{sec:newobserations} contains a gallery showing the four constituent 250 s DECam exposures, plus two images where isophotal contours were over-plotted to help identify coma extent for each of the first two exposures (Appendix \ref{sec:isophotalcontours}). 

We made use of three observatories for followup observations of \og{}: (1) NSF's OIR Labs DECam with $VR$ filter on the Blanco 4~m telescope at the Cerro Tololo Inter-American Observatory in Chile (2) WB4800-7800 filtered imaging with the Magellan 6.5~m Walter Baade  Telescope equipped with the Inamori-Magellan Areal Camera \& Spectrograph (IMACS) at the Las Campanas Observatory on Cerro Manqui, Chile, and (3) $g$, $r$, and $i$ filter images taken with the Large Monolithic Imager (LMI) at the Lowell Observatory 4.3~m Discovery Channel Telescope (DCT) in Arizona, USA.  Galleries showing our Magellan images and DCT images are shown in Appendices \ref{sec:magellanobservations} and \ref{sec:dctobservations}, respectively. A log of observations is provided in Appendix \ref{sec:observationdetails}. Astrometric calibration was performed using the \textit{astrometry.net} \citep{lang2010astrometrynet} and/or \textit{PhotometryPipeline} \citep{Mommert2017photometrypipeline} software packages.

\section{Simulating Dynamical Lifetime}
\label{sec:dynamicallifetimesimulation}
Determining the total mass loss possible for different volatiles requires knowledge of the dynamical lifetime of \og{} in the Centaur region (where both perihelion distance and semi-major axis are between 5 and 30~au). To this end we made use of the REBOUND $N$-body integrator to model the orbits of \og{} and giant planets Jupiter, Saturn, Uranus, and Neptune \citep{10.1093/mnras/stz769}. We also carried out 25 simulations of \og{}, each with an orbital clone derived from the orbital uncertainties published by the Minor Planet Center. From these dynamical integrations, we found that the lifetime of \og{} spans the range of 13,000 to 1.8 million years, roughly in agreement with prior work \citep{2019ApJ...880...71L}.

\section{Sublimation Modeling}
\label{sec:sublimationmodeling}

In order to better assess potential processes responsible for \og{} activity, we computed equilibrium temperatures and modeled mass loss rates for seven astrophysically relevant ices: ammonia (NH$_3$), carbon dioxide (CO$_2$), carbon monoxide (CO), methane (CH$_4$), methanol (CH$_3$OH), nitrogen (N$_2$) and water (H$_2$O). 

Object distance is the primary factor in determining potential ice sublimation effects. We began with a simple sublimation model \citep{Hsieh2015MBCsPS1} well-suited to gaining broad insight into the observed activity from \og{}; we expanded the procedure to apply more generally to other volatile ices. As we do not know the composition of \og{} we cannot make use of a more comprehensive model which includes effects of, for example, porosity, tortuosity, or crystal structure \citep{WOS:000257834000059}. Moreover, \og{} is undoubtedly not composed of a single ice, and mixtures of ices can exhibit behavior uncharacteristic for any lone constituent \citep{2000Icar..148..340G}. For the limiting case of an inert gray body orbiting at a distance $R$ from the Sun (measured in au)

\begin{equation}
    \frac{F_\odot}{R^2}(1-A)=\chi \epsilon \sigma T_\mathrm{eq}^4
\end{equation}

\noindent where the fiducial solar flux $F_\odot$ is 1360 W $/$ m$^{2}$, $A$ is the Bond albedo (we choose 0.1 as representative for Centaurs \citep{Peixinho2020CentaursReview}), $\epsilon$ is the infrared emissivity of the ice (set here as 0.9), $T_\mathrm{eq}$ is the equilibrium temperature of the body, and $\sigma$ is the Stefan-Boltzmann constant ($5.670\times10^{-8}$ W$/$m$^{2}\cdot$K$^{4}$). Here $\chi$ is a factor that describes the rotational and axial tilt effects on how much flux is received from the Sun: $\chi=1$ indicates the maximum heating scenario where the body is a ``slab'' facing the Sun at all times; $\chi=\pi$ describes a body that rotates quickly with no axial tilt with respect to the Sun; and $\chi=4$, which we adopt here, is used for a fast-rotating (on the order of a few hours) isothermal body in thermodynamic equilibrium;
here ``fast-rotating'' means that the rotation period of the object is short compared to the thermal wave propagation time \citep{WOS:000257834000059,Hsieh2015MBCsPS1}.

We next consider an energy balance that incorporates sublimation in addition to blackbody radiation \citep{Hsieh2015MBCsPS1}:

\begin{equation}
    \frac{F_\odot}{R^2} (1-A) = \chi\left[\epsilon\sigma T^4 + L f_\mathrm{D}\dot{m}_\mathrm{S}(T)\right]
    \label{eq:sublimationfull}
\end{equation}

\noindent where $f_\mathrm{D}$ is the ``diffusion barrier factor''  which describes how much emission is blocked by overlaying material (e.g., regolith), and $L$ the latent heat of sublimation. The mass loss rate $\dot{m}_\mathrm{S}(T)$ is given by

\begin{equation}
    \dot{m} = P_\mathrm{v}(T)\sqrt{\frac{\mu}{2\pi k T}}
\end{equation}
\noindent with $\mu$ the SI mass of one molecule, and $k$ the Boltzmann constant of $1.38069\times10^{-23}$ J$/$K. The vapor pressure (in Pa) of the substance can be related to temperature by the Clausius-Clapeyron relationship

\begin{equation}
    P_\mathrm{v}(T) = e_\mathrm{S} \exp \left[\frac{\Delta H_\mathrm{subl}}{R_\mathrm{g}}\left(\frac{1}{T_\mathrm{triple}}-\frac{1}{T}\right)\right]
\end{equation}

\noindent in which $e_\mathrm{S}$ is the saturation vapor pressure (in Pa) of the substance at the triple point temperature $T_\mathrm{triple}$, $\Delta H_\mathrm{subl}$ is the heat of sublimation of the substance (in kJ$/$mol), and $R_\mathrm{g}$ is the ideal gas constant ($\rm 8.341~J/mol\cdot K$).

Solving Equation \ref{eq:sublimationfull} for heliocentric distance $R$ (in au) yields

\begin{equation}
    R(T) = \sqrt{\frac{F_\odot (1-A)}{\chi\left[\epsilon\sigma T^4 + L f_\mathrm{D}\dot{m}_\mathrm{S}(T)\right]}}.
\end{equation}

Energy of sublimation values \citep{Luna2014AstroIces} and triple-point temperatures and pressures \citep{WOS:000273099100041} were incorporated as needed. 
To validate our model we computed the mass loss rate for (2060)~Chiron assuming $\chi=4$, an albedo of $0.057$, a diameter of 206 km, and an orbit ranging from 8.47 au at perihelion to 18.87~au at aphelion. Our (2060)~Chiron model validation results were in rough agreement with the 0.5 to 20~kg/s mass loss rate reported by \citet{womack2017co}.

\begin{figure*}
    \centering
    \includegraphics[width=0.85\linewidth]{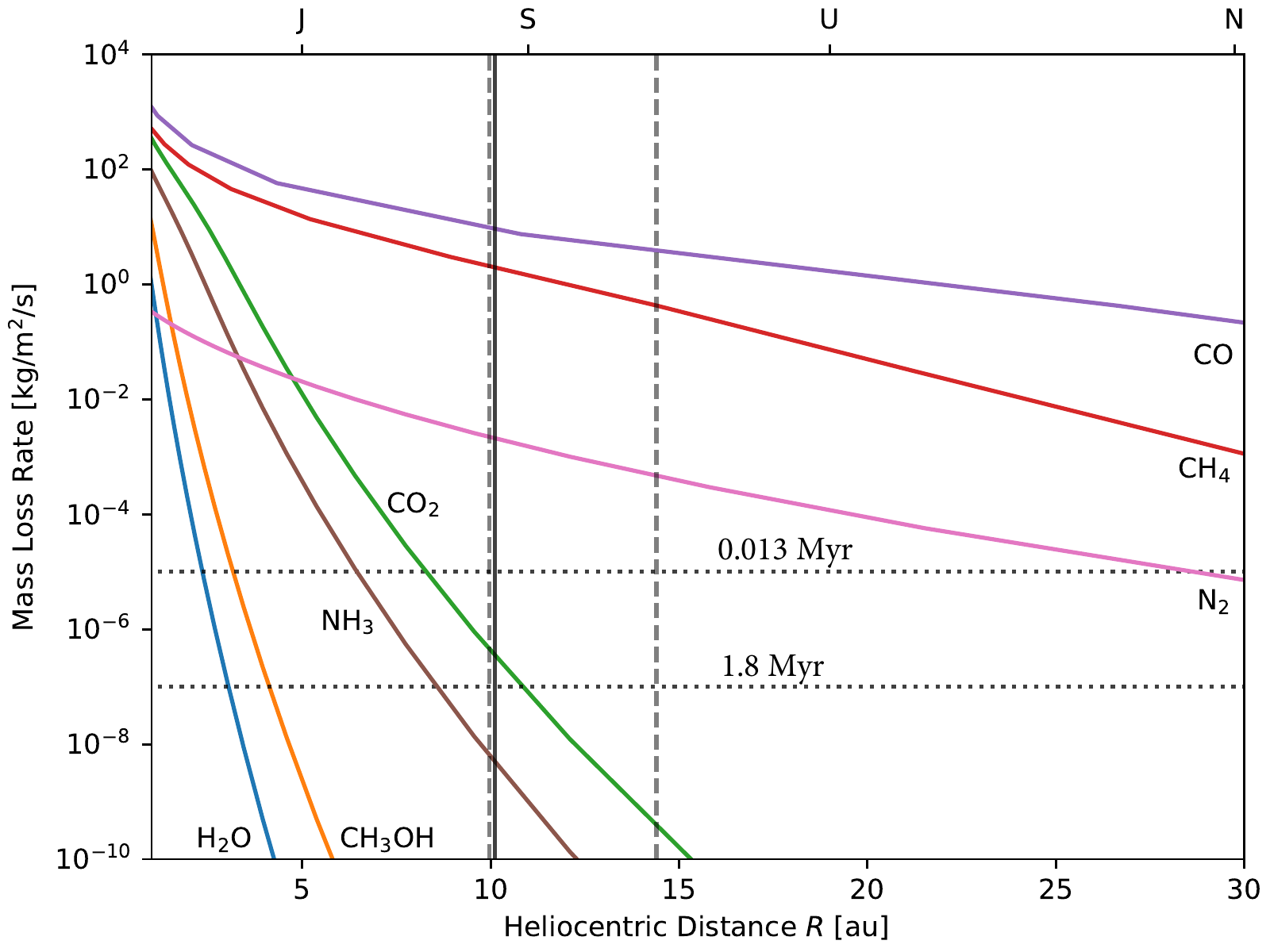}
    \caption{Mass loss rates for seven different astrophysically relevant ices on an isothermal ($\chi=4$) body; water (H$_2$O) and methanol (CH$_3$OH) ices have been detected on Centaurs. Orbital distances of Jupiter, Saturn, Uranus and Neptune are indicated about the top axis. The current 10.11~au heliocentric distance of \og{} is indicated by a vertical black bar, bracketed by perihelion (9.97 au) and aphelion (14.40~au) distances (leftmost and rightmost dashed vertical lines, respectively). Over the course of one orbit (between the vertical dashed lines), water and methanol never appreciably sublimate and carbon monoxide (CO), methane (CH$_4$), and molecular nitrogen (N$_2$) sublimate at high and relatively constant rates; we rule out all of these molecules as potential causes of activity. (The shallow slopes of CO, CH$_4$, and N$_2$ extend beyond 50~au [not shown] which informs us the mass loss would have begun long before \og{} became a Centaur.) However, over the course of one orbit the sublimation rates for CO$_2$ and NH$_3$ vary substantially, presumably producing  significant variation in visible activity. Order-of-magnitude estimates of mass loss rate upper limits for the dynamical lifetime of \og{} are shown as horizontal dotted lines. Only CO$_2$ and NH$_3$ have sublimation rates near these limits.}
    \label{fig:masslossrates}
\end{figure*}

We use our computed dynamical lifetime to circumstantially constrain the molecule(s) responsible for the sublimation of \og{}. Fig. \ref{fig:masslossrates} shows, over the orbit of \og{}, the mass loss rates for the different ices determined via modeling and validated through laboratory measurements. If \og{} has an albedo of 10\%, similar to that measured for other Centaurs (see review, \citealt{Peixinho2020CentaursReview}), then the body is about 20~km in diameter (see Section \ref{sec:absmaganddiameter}). Assuming a spherical body of low density in the range of 1~g/cm$^3$ to 3~g/cm$^3$ suggests a reasonable body mass of 4.2 to $12.6\times 10^{15}$~kg and a surface area of $3.1 \times 10^{8} \mbox{ m}^2$. Thus, the 13,000 to 1.8 Myr dynamical lifetime of \og{} suggests a maximum orbit-averaged mass loss rate in the range of $7.1\times10^{-7}$ to $\rm 3.3\times 10^{-5}\ kg/m^{2}/s$ (horizontal dashed lines in Figure \ref{fig:masslossrates}) before the body would be entirely lost due to sublimation. 


\section{Colors}
\label{sec:colors}
The archival data and our confirmation observations did not contain enough information to determine colors, so we obtained six 300~s exposures of \og{} in a $g$-$r$-$i$ filter sequence at the DCT (Section \ref{sec:observations}). We made use of the \textit{PhotometryPipeline} software package \citep{Mommert2017photometrypipeline} to automate astrometry using SCAMP \citep{Bertin2006SCAMP} which made use of the Vizier catalog service \citep{Ochsenbein2000Vizier} Gaia Data Release 2 catalog \citep{Gaia2018DR2}, and photometric image calibration using solar stars from the Sloan Digital Sky Survey Data Release 9 (SDSS-DR9) catalog \citep{SDSS2012DR9}. We carried out manual aperture photometry using the Aperture Photometry Tool \citep{Laher2012APT}.

\begin{figure}
    \centering
    \includegraphics[width=1.0\columnwidth]{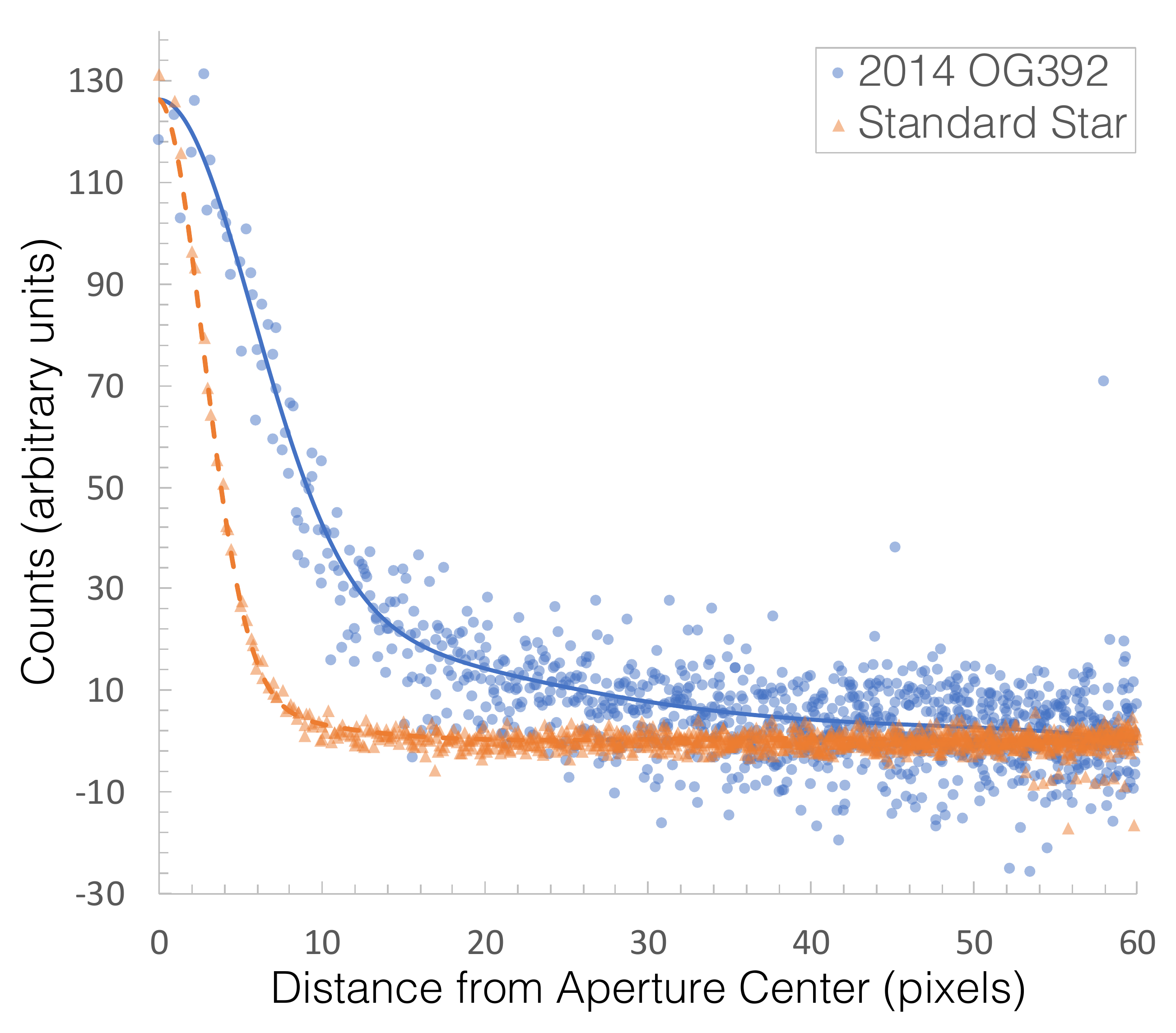}
    \caption{Surface brightness radial profiles of \og{} and a nearby SDSS-DR9 catalog Solar-type star (J004840.66-022335.6) are plotted along with a model fit for each object. After subtracting the background flux from the two profiles we normalized the standard star profile to the peak of the \og{} profile. The coma flux tapers from 125 counts to background (0 counts) at $\rho\simeq60$ pixels, or $4.3\times10^5$~km. We estimate there are $\sim5.8\times10^{17}$ particles in the coma assuming a grain radius of 1~mm; for a density of 1~g/cm$^3$ the total mass is $2.4\times10^{15}$~g. Data from our 300~s $g$-band exposure taken on UT 2019-12-30 2:29 using the Large Monolithic Imager on the Lowell Observatory 4.3~m Discovery Channel Telescope.} 
    \label{fig:sbrp}
\end{figure}

Prior to analysis we examined all thumbnail images showing activity emanating from \og{} to ensure no significant background sources were blended with the nucleus. To help us identify unseen contaminators we measured and modelled surface brightness radial profiles of \og{} (Figure \ref{fig:sbrp}) and a nearby Solar-type star, using Aperture Photometry Tool. The radial profile itself (i.e., not the model) was used to identify flux contribution by unseen background sources; we rejected images in which the nucleus or nearby coma was significantly contaminated. We note that we identified at least one background source within the coma in all of our images, although for color measurement we were able to use an aperture small enough (5 pixel radius) to exclude all resolvable background objects. 

We measured \og{} apparent magnitudes to be $g=21.99\pm0.018$, $r=21.19\pm0.016$, and $i=20.81\pm0.018$. We compared our colors of $g-r=0.80\pm0.024$ and $r-i=0.39\pm0.024$ to SDSS reported Solar colors of $g-r=0.44\pm0.02$ and $r-i=0.11\pm0.02$\footnote{\url{http://www.sdss.org/dr12/algorithms/ugrizvegasun}}. Centaur colors are often reported in Johnson $B$-$R$ colors (see e.g., \citealt{Tegler2016colors}), so we computed the $B-R$ color for \og{} via \cite{Jester2005SDSS} transformations. We found $B$-$R$ = \ogColor{} which is about one magnitude redder than the Sun, and red according to the classification system of \cite{Tegler2016colors} (see discussion in Section \ref{sec:discussion}).

\section{Absolute Magnitude and Diameter Estimation}
\label{sec:absmaganddiameter}

To gauge the overall spatial extent of the coma we examined the radial surface brightness profiles of \og{} and nearby Solar-type star J004840.66-022335.6 (see Section \ref{sec:colors}). We fit the profiles to the model

\begin{equation}
    S(r) = A + Br + Cr^2 + Dr^3 + Er^4 + F e^{-\frac{r^2}{2\sigma^2}}
\end{equation}

\noindent as described in \cite{2012PASP..124..579G}.

After subtracting the sky flux from each profile and each model we scaled the star to the peak flux of the \og{} radial profile. Figure \ref{fig:sbrp} shows the radial profiles and their corresponding models plotted; we estimate the coma returns to sky background flux levels at $\sim$60 pixels from the aperture center, thus the coma extent is $\sim4.3\times10^5$ km. The full width at half maximum (FWHM) of \og{} was $13.62\pm0.37$ pixels ($3.2\pm0.09\arcsec$), whereas the star FWHM was $6.05\pm0.05$ pixels ($1.45\pm0.012\arcsec$).

As reported in Section \ref{sec:sublimationmodeling}, the coma is likely present throughout the orbit of \og{}. As a result, prior absolute ($H$) magnitude estimates would have included the excess flux caused by the coma, as evinced in Figure \ref{fig:sbrp}. To estimate the absolute nuclear magnitude of \og{} we compared the ratio of the total (nucleus + coma) flux (blue line and circles, Figure \ref{fig:sbrp}) to the scaled stellar flux (orange line and triangles, Figure \ref{fig:sbrp}). We estimate the coma accounts for 0.75 and 1.1 magnitudes of the observed $r$-band and $g$-band flux, respectively, implying the nucleus apparent magnitudes are $m_r=21.9$ and $m_g=23.1$.

The absolute magnitude of an asteroid, $H$, is commonly used to estimate the size of small bodies in the Solar System . $H$ is defined as equal to the apparent $V$-band magnitude of an object observed at a heliocentric distance $R=1$ au, a geocentric distance $\Delta=1$ au, and a phase angle $\alpha=0^\circ$. Here we employ the International Astronomical Union defined \citep{1986IAUTB..19.....S} $H$-$G$ magnitude system approximated from \cite{1989aste.conf..524B}

\begin{equation}
	V = 5 \log \left(R \Delta\right) + H - 2.5 \log \left[\left(1-G\right)\Phi_1 + G \Phi_2\right]
	\label{eq:HG}
\end{equation}

\noindent where the phase function $\Phi$ is given by

\begin{equation}
	\Phi_i = \exp\left[-A_i\tan\left(\alpha/2\right)^{B_i}\right]; i=1,2
\end{equation}

\noindent with constants $A_1=3.33$, $A_2=1.87$, $B_1=0.63$, and $B_2=1.22$.

We make use of the relationships put forth by \cite{Jester2005SDSS} to derive Johnson $V=22.4$ from our $g$ and $r$ nuclear magnitudes. The JPL Horizons ephemerides service \citep{Giorgini1996Horizons} provided $G=0.150$ (the standard assumed slope for dark surfaces), $r=10.10$~au, $\Delta=10.01$~au, $\alpha=5.58^\circ$ for UT 2019-12-30. Via Equation~\ref{eq:HG} we find $H=11.3$, $0.5$ magnitudes fainter than reported by the Minor Planet Center and JPL Horizons.

\cite{1997Icar..126..450H} provide a convenient method to approximate object diameter $D$,

\begin{equation}
    D = \frac{1329}{\sqrt{G}}\times 10^{-H/5},
\end{equation}

\noindent which, for \og{}, gives $D\approx20$~km.




\section{Coma Dust Analysis}
\label{sec:dustproductionmodeling}

To facilitate comparing our \og{} dust-related metrics with other works we adopt the instrument and aperture-independent cometary dust production parameter described by \cite{1984AJ.....89..579A}. The metric, $Af\rho$ (units of cm), combines the mean albedo $A$ of ejecta grains within an aperture of radius $\rho$ (in cm), scaled by the filling factor $f$ (unitless) which describes how much of the aperture area ($\pi\rho^2$) is filled by $N$ grains of cross section area $\sigma$ (in cm$^2$)

\begin{equation}
    f = \frac{N(\rho)\sigma}{\pi\rho^2}.
    \label{eq:f}
\end{equation}

We measured $Af\rho$ (following the method outlined by \citealt{2019NatSR...9.5492S}) via

\begin{equation}
    A f \rho = 4 R^2 \Delta^2 10^{0.4(m_{\odot,F} - m_{\mathrm{OG},F})} \rho^{-1}
\end{equation}

\noindent where $R$ is the \og{} heliocentric distance in au, $\Delta$ is the geocentric distance of \og{} in cm, and, for filter $F$, $m_{\odot,F}$ and $m_{\mathrm{OG},F}$ are the magnitudes of the Sun and \og{}, respectively. For $m_{\odot,F}$ we made use of Solar apparent Vega magnitudes\footnote{\url{http://mips.as.arizona.edu/~cnaw/sun.html}} (see \citet{2018ApJS..236...47W} for details):

\begin{center}
    \begin{tabular}{cc}
        Filter & $m_{\odot,F}$\\
        \hline
        SDSS-$g$ & -26.34\\
        SDSS-$r$ & -27.04\\
        SDSS-$i$ & -27.38\\
    \end{tabular}
\end{center}

To estimate the number of particles $N$ within our measured $Af\rho$ we can substitute Equation \ref{eq:f} into the equality $Af\rho=Af\rho$

\begin{equation}
    Af\rho=A\frac{N(\rho)\sigma}{\pi\rho^2}\rho
\end{equation}

\noindent and solve for $N(\rho)$

\begin{equation}
    N(\rho) = Af\rho \frac{\pi\rho}{A\sigma}.
    \label{eq:Nrho}
\end{equation}

To quantify the total number of particles in the coma $N_\mathrm{tot}$ we can scale the aperture of Equation \ref{eq:Nrho} to the 60 pixel aperture containing the entire coma, $\rho_\mathrm{max}$,

\begin{equation}
    N(\rho_\mathrm{max}) = Af\rho \frac{\pi\rho_\mathrm{max}^2}{A \sigma \rho}.
    \label{eq:Nmax}
\end{equation}

\noindent Recall the quantity $Af\rho$, here, is a measured value, so the quantities $A\rho$ do not cancel in Equation \ref{eq:Nmax}.

Four of our observations, Images 15-18 (details in Appendix \ref{sec:observationdetails}) were suitable for directly measuring $Af\rho$. We found $Af\rho=487\pm 12$~cm with an aperture of $4.3\times10^5$~km. With the albedo adopted for our sublimation modeling ($A=0.1$) and a 1~mm radius grain, the coma around \og{} is composed of roughly $5.8\times10^{17}$ particles. Assuming a grain density of 1~g/cm$^3$ the total coma mass is $\sim2.4\times10^{15}$~g. 


\section{Discussion}
\label{sec:discussion}

The activity we observed spans more than two years which rules out impact-driven activity. We determined that the two ices previously detected on Centaurs, water and methanol, would not appreciably sublimate at any point in \og{}'s orbit and so should still be present in solid form on the surface (Figure \ref{fig:masslossrates}). Moreover, CO, N$_2$ and CH$_4$ are highly volatile and sublimate at temperatures low enough that their supply is likely depleted, though reservoirs could still be trapped below the surface. We reiterate our model encompasses single-species ices subjected to the thermodynamic conditions outlined in Section \ref{sec:sublimationmodeling}; heterogeneous ice environments may alter sublimation chemistry (see e.g., \citealt{2000Icar..148..340G}), as can single-species state transitions (e.g., energy released during crystallization of amorphous water ice; see e.g., \citealt{Jewitt2009TheActiveCentaurs}).

We find that the molecule(s) most likely to drive the observed activity is either $\rm CO_2$ and/or possibly $\rm NH_3$. Neither would have sublimated appreciably at Kuiper Belt distances prior to \og{} becoming a Centaur. Interestingly, both of these substances sublimate at rates that vary by over two orders of magnitude over the course of a \og{} orbit, peaking at perihelion. As a result we predict \og{} will become less active post-perihelion. This further implies that all other active Centaurs should follow this trend, with peak sublimation near perihelion and a significant drop in outgassing for most of their orbits.

We determined \og{} is at present roughly one magnitude redder than the Sun at visible wavelengths. However, we were only able to obtain two images in each filter, so uncertainty could be improved upon with additional observations. Our color measurements inexorably included the coma; future observations during a quiescent period (should one exist) would allow for color measurements of the bare nucleus. We did, however, attempt to better estimate the $H$ magnitude by subtracting the coma measured in the radial surface brightness profiles. We found \og{} has $H\approx11.3$, 0.5 magnitudes fainter than previously reported. The $H$ magnitude implies a radius of about 20~km when assuming a slope parameter $G=0.15$ as is typical for a dark surface. 

In our images of \og{} background sources were typically present in the coma and/or blended with the nucleus, but from four images we were able to directly measure dust properties. Assuming a 10\% albedo and a grain radius of 1~mm we estimate the coma contains roughly $5.8\times10^{17}$ particles. If the grain density is 1~g/cm$^3$, the total mass is $\sim2.4\times10^{15}$~g, or $\sim0.01$\% the total mass of \og{}. If the coma mass is indeed of this scale, \og{} must be eroding very quickly, undergoing new activity, or the ejecta is accumulating faster than it is escaping. Our measured $Af\rho$ of $487\pm12$~cm is comparable to other Centaurs active at the same orbital distance as \og{}: C/2011~P2~(PANSTARRS) with $Af\rho=161\pm 4$~cm at $\sim 9$~au \citep{2017A&A...597A..59M}, and for 166P~(NEAT) $Af\rho=288\pm 19$~cm at $\sim 12$~au \citep{2015MNRAS.454.3635S}. 


Centaurs are sometimes classified as either gray or red depending on whether the object has a $B$-$R$ color closer to $\sim$1.2 or $\sim$1.7, respectively (see \citealt{2008ssbn.book..105T} and \citealt{Peixinho2020CentaursReview} reviews for in-depth discussions). We find our derived $B$-$R$ color of \ogColor{} consistent with the red classification. Notably both molecules we find viable for sublimation are spectrally neutral in visible wavelengths so the reddening agent is as yet unidentified. 
\og{} will remain observable through February 2020 and will again be observable beginning around August 2020. We anticipate imaging and spectroscopy will yield further insight into the nature of these rare objects. We wish to emphasize further lab work is needed to characterize sublimation processes of volatiles under low pressure and temperature regimes.

\section{Acknowledgments}

We thank Dr.\ Mark Jesus Mendoza Magbanua (University of California San Francisco) for his frequent and timely feedback on the project. JKK acknowledges support from Northern Arizona University through a startup award administered by TD Robinson. Mark Loeffler (NAU) and Patrick Tribbett (NAU) helped us interpret lab results. Stephen Tegler (NAU) provided extensive insight into the nuances of Centaur colors. Kazuo Kinoshita provided cometary elements which saved us considerable time and energy. The authors express their gratitude to Mike Gowanlock (NAU), Cristina Thomas (NAU), and the Trilling Research Group (NAU), all of whom provided insights which substantially enhanced this work. Thank you to Stephen Kane (University of California Riverside), Dawn Gelino (NASA Exoplanet Science Institute at California Institute of Technology), and Jonathan Fortney (University of California Santa Cruz) for encouraging the authors to pursue this work. The unparalleled support provided by Monsoon cluster administrator Christopher Coffey (NAU) and his High Performance Computing Support team greatly facilitated the work presented here.

\ This material is based upon work supported by the National Science Foundation Graduate Research Fellowship Program under Grant No.\ 2018258765. Any opinions, findings, and conclusions or recommendations expressed in this material are those of the author(s) and do not necessarily reflect the views of the National Science Foundation.

\ Computational analyses were carried out on Northern Arizona University's Monsoon computing cluster, funded by Arizona's Technology and Research Initiative Fund.
\ This work was made possible in part through the State of Arizona Technology and Research Initiative Program.

\ This research used the facilities of the Canadian Astronomy Data Centre operated by the National Research Council of Canada with the support of the Canadian Space Agency. We also employed their Solar System Object Search \citep{2012PASP..124..579G}.
\ This research has made use of data and/or services provided by the International Astronomical Union's Minor Planet Center.
\ This research has made use of NASA's Astrophysics Data System.
\ This research has made use of the The Institut de M\'ecanique C\'eleste et de Calcul des \'Eph\'em\'erides (IMCCE) SkyBoT Virtual Observatory tool \citep{Berthier:2006tn}.
\ Simulations in this paper made use of the REBOUND code which is freely available at http://github.com/hannorein/rebound.
\ This work made use of the {FTOOLS} software package hosted by the NASA Goddard Flight Center High Energy Astrophysics Science Archive Research Center.
\ This research has made use of SAO ImageDS9, developed by Smithsonian Astrophysical Observatory \citep{2003ASPC..295..489J}. This work made use of the Lowell Observatory Asteroid Orbit Database \textit{astorbDB} \citep{AstOrb2019EPSCDPS}.
\ This work made use of the \textit{astropy} software package \citep{AstroPy2013}.

\ This project used data obtained with the Dark Energy Camera (DECam), which was constructed by the Dark Energy Survey (DES) collaboration. Funding for the DES Projects has been provided by the U.S. Department of Energy, the U.S. National Science Foundation, the Ministry of Science and Education of Spain, the Science and Technology Facilities Council of the United Kingdom, the Higher Education Funding Council for England, the National Center for Supercomputing Applications at the University of Illinois at Urbana-Champaign, the Kavli Institute of Cosmological Physics at the University of Chicago, Center for Cosmology and Astro-Particle Physics at the Ohio State University, the Mitchell Institute for Fundamental Physics and Astronomy at Texas A\&M University, Financiadora de Estudos e Projetos, Funda\c{c}\~{a}o Carlos Chagas Filho de Amparo, Financiadora de Estudos e Projetos, Funda\c{c}\~ao Carlos Chagas Filho de Amparo \`{a} Pesquisa do Estado do Rio de Janeiro, Conselho Nacional de Desenvolvimento Cient\'{i}fico e Tecnol\'{o}gico and the Minist\'{e}rio da Ci\^{e}ncia, Tecnologia e Inova\c{c}\~{a}o, the Deutsche Forschungsgemeinschaft and the Collaborating Institutions in the Dark Energy Survey. The Collaborating Institutions are Argonne National Laboratory, the University of California at Santa Cruz, the University of Cambridge, Centro de Investigaciones Energ\'{e}ticas, Medioambientales y Tecnol\'{o}gicas–Madrid, the University of Chicago, University College London, the DES-Brazil Consortium, the University of Edinburgh, the Eidgen\"ossische Technische Hochschule (ETH) Z\"urich, Fermi National Accelerator Laboratory, the University of Illinois at Urbana-Champaign, the Institut de Ci\`{e}ncies de l'Espai (IEEC/CSIC), the Institut de Física d'Altes Energies, Lawrence Berkeley National Laboratory, the Ludwig-Maximilians Universit\"{a}t M\"{u}nchen and the associated Excellence Cluster Universe, the University of Michigan, NSF’s National Optical-Infrared Astronomy Research Laboratory, the University of Nottingham, the Ohio State University, the University of Pennsylvania, the University of Portsmouth, SLAC National Accelerator Laboratory, Stanford University, the University of Sussex, and Texas A\&M University.

\ This work is based in part on observations at Cerro Tololo Inter-American Observatory, National Optical Astronomy Observatory (Prop. IDs 2019A-0337, PI: Trilling; 2014B-0404, PI: Schlegel), which is operated by the Association of Universities for Research in Astronomy (AURA) under a cooperative agreement with the National Science Foundation.

These results made use of the Discovery Channel Telescope at Lowell Observatory. Lowell is a private, non-profit institution dedicated to astrophysical research and public appreciation of astronomy and operates the DCT in partnership with Boston University, the University of Maryland, the University of Toledo, Northern Arizona University and Yale University. The Large Monolithic Imager was built by Lowell Observatory using funds provided by the National Science Foundation (AST-1005313). This paper includes data gathered with the 6.5 meter Magellan Telescopes located at Las Campanas Observatory, Chile.


\begin{thebibliography}{}
\expandafter\ifx\csname natexlab\endcsname\relax\def\natexlab#1{#1}\fi
\providecommand{\url}[1]{\href{#1}{#1}}
\providecommand{\dodoi}[1]{doi:~\href{http://doi.org/#1}{\nolinkurl{#1}}}
\providecommand{\doeprint}[1]{\href{http://ascl.net/#1}{\nolinkurl{http://ascl.net/#1}}}
\providecommand{\doarXiv}[1]{\href{https://arxiv.org/abs/#1}{\nolinkurl{https://arxiv.org/abs/#1}}}

\bibitem[{{A'Hearn} {et~al.}(1984){A'Hearn}, {Schleicher}, {Millis}, {Feldman},
  \& {Thompson}}]{1984AJ.....89..579A}
{A'Hearn}, M.~F., {Schleicher}, D.~G., {Millis}, R.~L., {Feldman}, P.~D., \&
  {Thompson}, D.~T. 1984, \aj, 89, 579, \dodoi{10.1086/113552}

\bibitem[{{Ahn} {et~al.}(2012){Ahn}, {Alexandroff}, {Allende Prieto},
  {Anderson}, {Anderton}, {Andrews}, {Aubourg}, {Bailey}, {Balbinot}, {Barnes},
  {Bautista}, {Beers}, {Beifiori}, {Berlind}, {Bhardwaj}, {Bizyaev}, {Blake},
  {Blanton}, {Blomqvist}, {Bochanski}, {Bolton}, {Borde}, {Bovy}, {Brandt},
  {Brinkmann}, {Brown}, {Brownstein}, {Bundy}, {Busca}, {Carithers}, {Carnero},
  {Carr}, {Casetti-Dinescu}, {Chen}, {Chiappini}, {Comparat}, {Connolly},
  {Crepp}, {Cristiani}, {Croft}, {Cuesta}, {da Costa}, {Davenport}, {Dawson},
  {de Putter}, {De Lee}, {Delubac}, {Dhital}, {Ealet}, {Ebelke}, {Edmondson},
  {Eisenstein}, {Escoffier}, {Esposito}, {Evans}, {Fan}, {Femen{\'\i}a
  Castell{\'a}}, {Fern{\'a}ndez Alvar}, {Ferreira}, {Filiz Ak}, {Finley},
  {Fleming}, {Font-Ribera}, {Frinchaboy}, {Garc{\'\i}a-Hern{\'a}ndez},
  {Garc{\'\i}a P{\'e}rez}, {Ge}, {G{\'e}nova-Santos}, {Gillespie}, {Girardi},
  {Gonz{\'a}lez Hern{\'a}ndez}, {Grebel}, {Gunn}, {Guo}, {Haggard}, {Hamilton},
  {Harris}, {Hawley}, {Hearty}, {Ho}, {Hogg}, {Holtzman}, {Honscheid},
  {Huehnerhoff}, {Ivans}, {Ivezi{\'c}}, {Jacobson}, {Jiang}, {Johansson},
  {Johnson}, {Kauffmann}, {Kirkby}, {Kirkpatrick}, {Klaene}, {Knapp}, {Kneib},
  {Le Goff}, {Leauthaud}, {Lee}, {Lee}, {Long}, {Loomis}, {Lucatello},
  {Lundgren}, {Lupton}, {Ma}, {Ma}, {MacDonald}, {Mack}, {Mahadevan}, {Maia},
  {Majewski}, {Makler}, {Malanushenko}, {Malanushenko}, {Manchado},
  {Mandelbaum}, {Manera}, {Maraston}, {Margala}, {Martell}, {McBride},
  {McGreer}, {McMahon}, {M{\'e}nard}, {Meszaros}, {Miralda-Escud{\'e}},
  {Montero-Dorta}, {Montesano}, {Morrison}, {Muna}, {Munn}, {Murayama},
  {Myers}, {Neto}, {Nguyen}, {Nichol}, {Nidever}, {Noterdaeme}, {Nuza}, {Ogand
  o}, {Olmstead}, {Oravetz}, {Owen}, {Padmanabhan}, {Palanque-Delabrouille},
  {Pan}, {Parejko}, {Parihar}, {P{\^a}ris}, {Pattarakijwanich}, {Pepper},
  {Percival}, {P{\'e}rez-Fournon}, {P{\'e}rez-R{\`a}fols}, {Petitjean},
  {Pforr}, {Pieri}, {Pinsonneault}, {Porto de Mello}, {Prada}, {Price-Whelan},
  {Raddick}, {Rebolo}, {Rich}, {Richards}, {Robin}, {Rocha-Pinto}, {Rockosi},
  {Roe}, {Ross}, {Ross}, {Rossi}, {Rubi{\~n}o-Martin}, {Samushia}, {Sanchez
  Almeida}, {S{\'a}nchez}, {Santiago}, {Sayres}, {Schlegel}, {Schlesinger},
  {Schmidt}, {Schneider}, {Schultheis}, {Schwope}, {Sc{\'o}ccola}, {Seljak},
  {Sheldon}, {Shen}, {Shu}, {Simmerer}, {Simmons}, {Skibba}, {Skrutskie},
  {Slosar}, {Sobreira}, {Sobeck}, {Stassun}, {Steele}, {Steinmetz}, {Strauss},
  {Streblyanska}, {Suzuki}, {Swanson}, {Tal}, {Thakar}, {Thomas}, {Thompson},
  {Tinker}, {Tojeiro}, {Tremonti}, {Vargas Maga{\~n}a}, {Verde}, {Viel},
  {Vikas}, {Vogt}, {Wake}, {Wang}, {Weaver}, {Weinberg}, {Weiner}, {West},
  {White}, {Wilson}, {Wisniewski}, {Wood-Vasey}, {Yanny}, {Y{\`e}che}, {York},
  {Zamora}, {Zasowski}, {Zehavi}, {Zhao}, {Zheng}, {Zhu}, \&
  {Zinn}}]{SDSS2012DR9}
{Ahn}, C.~P., {Alexandroff}, R., {Allende Prieto}, C., {et~al.} 2012, \apjs,
  203, 21, \dodoi{10.1088/0067-0049/203/2/21}

\bibitem[{{Astropy Collaboration} {et~al.}(2013){Astropy Collaboration},
  {Robitaille}, {Tollerud}, {Greenfield}, {Droettboom}, {Bray}, {Aldcroft},
  {Davis}, {Ginsburg}, {Price-Whelan}, {Kerzendorf}, {Conley}, {Crighton},
  {Barbary}, {Muna}, {Ferguson}, {Grollier}, {Parikh}, {Nair}, {Unther},
  {Deil}, {Woillez}, {Conseil}, {Kramer}, {Turner}, {Singer}, {Fox}, {Weaver},
  {Zabalza}, {Edwards}, {Azalee Bostroem}, {Burke}, {Casey}, {Crawford},
  {Dencheva}, {Ely}, {Jenness}, {Labrie}, {Lim}, {Pierfederici}, {Pontzen},
  {Ptak}, {Refsdal}, {Servillat}, \& {Streicher}}]{AstroPy2013}
{Astropy Collaboration}, {Robitaille}, T.~P., {Tollerud}, E.~J., {et~al.} 2013,
  \aap, 558, A33, \dodoi{10.1051/0004-6361/201322068}

\bibitem[{Bacci {et~al.}(2015)Bacci, Tesi, Fagioli, Bressi, Scotti, Gilmore,
  Kilmartin, Dupouy, de~Vanssay, Hidas, Gibson, Goggia, Primak, Schultz,
  Willman, Chambers, Chastel, Denneau, Flewelling, Huber, Lilly, Magnier,
  Wainscoat, Waters, Weryk, Oreshko, Losse, Sato, Maury, Bosch, Noel, \&
  Williams}]{Bacci2015P2015M2MPEC}
Bacci, P., Tesi, L., Fagioli, G., {et~al.} 2015, Minor Planet Electronic
  Circulars, 2015-N46

\bibitem[{{Berthier} {et~al.}(2006){Berthier}, {Vachier}, {Thuillot},
  {Fernique}, {Ochsenbein}, {Genova}, {Lainey}, \& {Arlot}}]{Berthier:2006tn}
{Berthier}, J., {Vachier}, F., {Thuillot}, W., {et~al.} 2006, Astronomical
  Society of the Pacific Conference Series, Vol. 351, {SkyBoT, a new VO service
  to identify Solar System objects}, ed. C.~{Gabriel}, C.~{Arviset}, D.~{Ponz},
  \& S.~{Enrique}, 367

\bibitem[{{Bertin}(2006)}]{Bertin2006SCAMP}
{Bertin}, E. 2006, Astronomical Society of the Pacific Conference Series, Vol.
  351, {Automatic Astrometric and Photometric Calibration with SCAMP}, ed.
  C.~{Gabriel}, C.~{Arviset}, D.~{Ponz}, \& S.~{Enrique}, 112

\bibitem[{{Bowell} {et~al.}(1989){Bowell}, {Hapke}, {Domingue}, {Lumme},
  {Peltoniemi}, \& {Harris}}]{1989aste.conf..524B}
{Bowell}, E., {Hapke}, B., {Domingue}, D., {et~al.} 1989, in Asteroids II, ed.
  R.~P. {Binzel}, T.~{Gehrels}, \& M.~S. {Matthews}, 524--556

\bibitem[{Busarev {et~al.}(2018)Busarev, Makalkin, Vilas, Barabanov, \&
  Scherbina}]{2018Icar..304...83B}
Busarev, V.~V., Makalkin, A.~B., Vilas, F., Barabanov, S.~I., \& Scherbina,
  M.~P. 2018, Icarus, 304, 83, \dodoi{10.1016/j.icarus.2017.06.032}

\bibitem[{Chandler {et~al.}(2018)Chandler, Curtis, Mommert, Sheppard, \&
  Trujillo}]{Chandler2018SAFARI}
Chandler, C.~O., Curtis, A.~M., Mommert, M., Sheppard, S.~S., \& Trujillo,
  C.~A. 2018, Publications of the Astronomical Society of the Pacific, 130,
  114502, \dodoi{10.1088/1538-3873/aad03d}

\bibitem[{Choi {et~al.}(2006)Choi, Weissman, \&
  Polishook}]{2006IAUC.8656....2C}
Choi, Y.~J., Weissman, P.~R., \& Polishook, D. 2006, IAU Circ., 8656, 2

\bibitem[{Cunningham(1950)}]{r02043}
Cunningham, L.~E. 1950, IAU Circ., 1250, 3

\bibitem[{{Fray} \& {Schmitt}(2009)}]{WOS:000273099100041}
{Fray}, N., \& {Schmitt}, B. 2009, \planss, 57, 2053,
  \dodoi{10.1016/j.pss.2009.09.011}

\bibitem[{{Gaia Collaboration} {et~al.}(2018){Gaia Collaboration}, {Brown},
  {Vallenari}, {Prusti}, {de Bruijne}, {Babusiaux}, {Bailer-Jones}, {Biermann},
  {Evans}, {Eyer}, {Jansen}, {Jordi}, {Klioner}, {Lammers}, {Lindegren},
  {Luri}, {Mignard}, {Panem}, {Pourbaix}, {Randich}, {Sartoretti}, {Siddiqui},
  {Soubiran}, {van Leeuwen}, {Walton}, {Arenou}, {Bastian}, {Cropper},
  {Drimmel}, {Katz}, {Lattanzi}, {Bakker}, {Cacciari}, {Casta{\~n}eda},
  {Chaoul}, {Cheek}, {De Angeli}, {Fabricius}, {Guerra}, {Holl}, {Masana},
  {Messineo}, {Mowlavi}, {Nienartowicz}, {Panuzzo}, {Portell}, {Riello},
  {Seabroke}, {Tanga}, {Th{\'e}venin}, {Gracia-Abril}, {Comoretto},
  {Garcia-Reinaldos}, {Teyssier}, {Altmann}, {Andrae}, {Audard},
  {Bellas-Velidis}, {Benson}, {Berthier}, {Blomme}, {Burgess}, {Busso},
  {Carry}, {Cellino}, {Clementini}, {Clotet}, {Creevey}, {Davidson}, {De
  Ridder}, {Delchambre}, {Dell'Oro}, {Ducourant},
  {Fern{\'a}ndez-Hern{\'a}ndez}, {Fouesneau}, {Fr{\'e}mat}, {Galluccio},
  {Garc{\'\i}a-Torres}, {Gonz{\'a}lez-N{\'u}{\~n}ez}, {Gonz{\'a}lez-Vidal},
  {Gosset}, {Guy}, {Halbwachs}, {Hambly}, {Harrison}, {Hern{\'a}ndez},
  {Hestroffer}, {Hodgkin}, {Hutton}, {Jasniewicz}, {Jean-Antoine-Piccolo},
  {Jordan}, {Korn}, {Krone-Martins}, {Lanzafame}, {Lebzelter}, {L{\"o}ffler},
  {Manteiga}, {Marrese}, {Mart{\'\i}n-Fleitas}, {Moitinho}, {Mora}, {Muinonen},
  {Osinde}, {Pancino}, {Pauwels}, {Petit}, {Recio-Blanco}, {Richards},
  {Rimoldini}, {Robin}, {Sarro}, {Siopis}, {Smith}, {Sozzetti}, {S{\"u}veges},
  {Torra}, {van Reeven}, {Abbas}, {Abreu Aramburu}, {Accart}, {Aerts},
  {Altavilla}, {{\'A}lvarez}, {Alvarez}, {Alves}, {Anderson}, {Andrei},
  {Anglada Varela}, {Antiche}, {Antoja}, {Arcay}, {Astraatmadja}, {Bach},
  {Baker}, {Balaguer-N{\'u}{\~n}ez}, {Balm}, {Barache}, {Barata}, {Barbato},
  {Barblan}, {Barklem}, {Barrado}, {Barros}, {Barstow}, {Bartholom{\'e}
  Mu{\~n}oz}, {Bassilana}, {Becciani}, {Bellazzini}, {Berihuete}, {Bertone},
  {Bianchi}, {Bienaym{\'e}}, {Blanco-Cuaresma}, {Boch}, {Boeche}, {Bombrun},
  {Borrachero}, {Bossini}, {Bouquillon}, {Bourda}, {Bragaglia}, {Bramante},
  {Breddels}, {Bressan}, {Brouillet}, {Br{\"u}semeister}, {Brugaletta},
  {Bucciarelli}, {Burlacu}, {Busonero}, {Butkevich}, {Buzzi}, {Caffau},
  {Cancelliere}, {Cannizzaro}, {Cantat-Gaudin}, {Carballo}, {Carlucci},
  {Carrasco}, {Casamiquela}, {Castellani}, {Castro-Ginard}, {Charlot},
  {Chemin}, {Chiavassa}, {Cocozza}, {Costigan}, {Cowell}, {Crifo}, {Crosta},
  {Crowley}, {Cuypers}, {Dafonte}, {Damerdji}, {Dapergolas}, {David}, {David},
  {de Laverny}, {De Luise}, {De March}, {de Martino}, {de Souza}, {de Torres},
  {Debosscher}, {del Pozo}, {Delbo}, {Delgado}, {Delgado}, {Di Matteo},
  {Diakite}, {Diener}, {Distefano}, {Dolding}, {Drazinos}, {Dur{\'a}n},
  {Edvardsson}, {Enke}, {Eriksson}, {Esquej}, {Eynard Bontemps}, {Fabre},
  {Fabrizio}, {Faigler}, {Falc{\~a}o}, {Farr{\`a}s Casas}, {Federici},
  {Fedorets}, {Fernique}, {Figueras}, {Filippi}, {Findeisen}, {Fonti},
  {Fraile}, {Fraser}, {Fr{\'e}zouls}, {Gai}, {Galleti}, {Garabato},
  {Garc{\'\i}a-Sedano}, {Garofalo}, {Garralda}, {Gavel}, {Gavras}, {Gerssen},
  {Geyer}, {Giacobbe}, {Gilmore}, {Girona}, {Giuffrida}, {Glass}, {Gomes},
  {Granvik}, {Gueguen}, {Guerrier}, {Guiraud}, {Guti{\'e}rrez-S{\'a}nchez},
  {Haigron}, {Hatzidimitriou}, {Hauser}, {Haywood}, {Heiter}, {Helmi}, {Heu},
  {Hilger}, {Hobbs}, {Hofmann}, {Holland}, {Huckle}, {Hypki}, {Icardi},
  {Jan{\ss}en}, {Jevardat de Fombelle}, {Jonker}, {Juh{\'a}sz}, {Julbe},
  {Karampelas}, {Kewley}, {Klar}, {Kochoska}, {Kohley}, {Kolenberg},
  {Kontizas}, {Kontizas}, {Koposov}, {Kordopatis}, {Kostrzewa-Rutkowska},
  {Koubsky}, {Lambert}, {Lanza}, {Lasne}, {Lavigne}, {Le Fustec}, {Le
  Poncin-Lafitte}, {Lebreton}, {Leccia}, {Leclerc}, {Lecoeur-Taibi},
  {Lenhardt}, {Leroux}, {Liao}, {Licata}, {Lindstr{\o}m}, {Lister}, {Livanou},
  {Lobel}, {L{\'o}pez}, {Managau}, {Mann}, {Mantelet}, {Marchal}, {Marchant},
  {Marconi}, {Marinoni}, {Marschalk{\'o}}, {Marshall}, {Martino}, {Marton},
  {Mary}, {Massari}, {Matijevi{\v{c}}}, {Mazeh}, {McMillan}, {Messina},
  {Michalik}, {Millar}, {Molina}, {Molinaro}, {Moln{\'a}r}, {Montegriffo},
  {Mor}, {Morbidelli}, {Morel}, {Morris}, {Mulone}, {Muraveva}, {Musella},
  {Nelemans}, {Nicastro}, {Noval}, {O'Mullane}, {Ord{\'e}novic},
  {Ord{\'o}{\~n}ez-Blanco}, {Osborne}, {Pagani}, {Pagano}, {Pailler},
  {Palacin}, {Palaversa}, {Panahi}, {Pawlak}, {Piersimoni}, {Pineau}, {Plachy},
  {Plum}, {Poggio}, {Poujoulet}, {Pr{\v{s}}a}, {Pulone}, {Racero}, {Ragaini},
  {Rambaux}, {Ramos-Lerate}, {Regibo}, {Reyl{\'e}}, {Riclet}, {Ripepi}, {Riva},
  {Rivard}, {Rixon}, {Roegiers}, {Roelens}, {Romero-G{\'o}mez}, {Rowell},
  {Royer}, {Ruiz-Dern}, {Sadowski}, {Sagrist{\`a} Sell{\'e}s}, {Sahlmann},
  {Salgado}, {Salguero}, {Sanna}, {Santana-Ros}, {Sarasso}, {Savietto},
  {Schultheis}, {Sciacca}, {Segol}, {Segovia}, {S{\'e}gransan}, {Shih},
  {Siltala}, {Silva}, {Smart}, {Smith}, {Solano}, {Solitro}, {Sordo}, {Soria
  Nieto}, {Souchay}, {Spagna}, {Spoto}, {Stampa}, {Steele},
  {Steidelm{\"u}ller}, {Stephenson}, {Stoev}, {Suess}, {Surdej}, {Szabados},
  {Szegedi-Elek}, {Tapiador}, {Taris}, {Tauran}, {Taylor}, {Teixeira},
  {Terrett}, {Teyssand ier}, {Thuillot}, {Titarenko}, {Torra Clotet}, {Turon},
  {Ulla}, {Utrilla}, {Uzzi}, {Vaillant}, {Valentini}, {Valette}, {van Elteren},
  {Van Hemelryck}, {van Leeuwen}, {Vaschetto}, {Vecchiato}, {Veljanoski},
  {Viala}, {Vicente}, {Vogt}, {von Essen}, {Voss}, {Votruba}, {Voutsinas},
  {Walmsley}, {Weiler}, {Wertz}, {Wevers}, {Wyrzykowski}, {Yoldas},
  {{\v{Z}}erjal}, {Ziaeepour}, {Zorec}, {Zschocke}, {Zucker}, {Zurbach}, \&
  {Zwitter}}]{Gaia2018DR2}
{Gaia Collaboration}, {Brown}, A.~G.~A., {Vallenari}, A., {et~al.} 2018, \aap,
  616, A1, \dodoi{10.1051/0004-6361/201833051}

\bibitem[{Gajdos {et~al.}(2005)Gajdos, Vilagi, Naves, Campas, McMillan, Durig,
  Yager, Ferrando, Gerashchenko, Ivashchenko, Kyrylenko, Lacruz, \&
  Marsden}]{Gajdos2005discP2005S2Skiff}
Gajdos, S., Vilagi, J., Naves, R., {et~al.} 2005, Minor Planet Electronic
  Circulars, 2005-T26

\bibitem[{Gibbs {et~al.}(2011{\natexlab{a}})Gibbs, Sarneczky, Scotti, Foglia,
  Vorobjov, Holmes, \& Williams}]{Gibbs2011P2011C2IAUC}
Gibbs, A.~R., Sarneczky, K., Scotti, J.~V., {et~al.} 2011{\natexlab{a}},
  International Astronomical Union Circular, 9199, 1

\bibitem[{Gibbs {et~al.}(2011{\natexlab{b}})Gibbs, Tornero, \&
  Williams}]{Gibbs2011C2011S1IAUC}
Gibbs, A.~R., Tornero, S.~F., \& Williams, G.~V. 2011{\natexlab{b}},
  International Astronomical Union Circular, 9234, 1

\bibitem[{{Giorgini} {et~al.}(1996){Giorgini}, {Yeomans}, {Chamberlin},
  {Chodas}, {Jacobson}, {Keesey}, {Lieske}, {Ostro}, {Standish}, \&
  {Wimberly}}]{Giorgini1996Horizons}
{Giorgini}, J.~D., {Yeomans}, D.~K., {Chamberlin}, A.~B., {et~al.} 1996, in
  AAS/Division for Planetary Sciences Meeting Abstracts \#28, AAS/Division for
  Planetary Sciences Meeting Abstracts, 25.04

\bibitem[{{Grundy} \& {Stansberry}(2000)}]{2000Icar..148..340G}
{Grundy}, W.~M., \& {Stansberry}, J.~A. 2000, Icarus, 148, 340,
  \dodoi{10.1006/icar.2000.6510}

\bibitem[{{Gwyn} {et~al.}(2012){Gwyn}, {Hill}, \&
  {Kavelaars}}]{2012PASP..124..579G}
{Gwyn}, S. D.~J., {Hill}, N., \& {Kavelaars}, J.~J. 2012, Publications of the
  Astronomical Society of the Pacific, 124, 579, \dodoi{10.1086/666462}

\bibitem[{{Harris} \& {Harris}(1997)}]{1997Icar..126..450H}
{Harris}, A.~W., \& {Harris}, A.~W. 1997, \icarus, 126, 450,
  \dodoi{10.1006/icar.1996.5664}

\bibitem[{Holvorcem {et~al.}(2013)Holvorcem, Schwartz, Sato, Stevens, \&
  Williams}]{Holvorcem2013C2013C2CBET}
Holvorcem, P.~R., Schwartz, M., Sato, H., Stevens, B.~L., \& Williams, G.~V.
  2013, Central Bureau Electronic Telegrams, 3417, 1

\bibitem[{{Horner} \& {Lykawka}(2010)}]{2010MNRAS.402...13H}
{Horner}, J., \& {Lykawka}, P.~S. 2010, \mnras, 402, 13,
  \dodoi{10.1111/j.1365-2966.2009.15702.x}

\bibitem[{{Hsieh} {et~al.}(2015){Hsieh}, {Denneau}, {Wainscoat},
  {Sch{\"o}rghofer}, {Bolin}, {Fitzsimmons}, {Jedicke}, {Kleyna}, {Micheli},
  {Vere{\v{s}}}, {Kaiser}, {Chambers}, {Burgett}, {Flewelling}, {Hodapp},
  {Magnier}, {Morgan}, {Price}, {Tonry}, \& {Waters}}]{Hsieh2015MBCsPS1}
{Hsieh}, H.~H., {Denneau}, L., {Wainscoat}, R.~J., {et~al.} 2015, \icarus, 248,
  289, \dodoi{10.1016/j.icarus.2014.10.031}

\bibitem[{{Jester} {et~al.}(2005){Jester}, {Schneider}, {Richards}, {Green},
  {Schmidt}, {Hall}, {Strauss}, {Vand en Berk}, {Stoughton}, {Gunn},
  {Brinkmann}, {Kent}, {Smith}, {Tucker}, \& {Yanny}}]{Jester2005SDSS}
{Jester}, S., {Schneider}, D.~P., {Richards}, G.~T., {et~al.} 2005, \aj, 130,
  873, \dodoi{10.1086/432466}

\bibitem[{{Jewitt}(2009)}]{Jewitt2009TheActiveCentaurs}
{Jewitt}, D. 2009, \aj, 137, 4296, \dodoi{10.1088/0004-6256/137/5/4296}

\bibitem[{{Joye} \& {Mandel}(2003)}]{2003ASPC..295..489J}
{Joye}, W.~A., \& {Mandel}, E. 2003, Astronomical Society of the Pacific
  Conference Series, Vol. 295, {New Features of SAOImage DS9}, ed. H.~E.
  {Payne}, R.~I. {Jedrzejewski}, \& R.~N. {Hook}, 489

\bibitem[{{Kowal} \& {Gehrels}(1977)}]{1977IAUC.3129....1K}
{Kowal}, C.~T., \& {Gehrels}, T. 1977, IAU Circular, 3129, 1

\bibitem[{Kowalski {et~al.}(2016)Kowalski, Birtwhistle, Sato, Sarneczky, Weryk,
  \& Williams}]{Kowalski2016C2016Q4activity}
Kowalski, R.~A., Birtwhistle, P., Sato, H., {et~al.} 2016, Central Bureau
  Electronic Telegrams, 4314

\bibitem[{Kusnirak \& Balam(2000)}]{2005IAUC.8552....2G}
Kusnirak, P., \& Balam, D. 2000, International Astronomical Union Circular,
  7368, 2

\bibitem[{{Laher} {et~al.}(2012){Laher}, {Gorjian}, {Rebull}, {Masci},
  {Fowler}, {Helou}, {Kulkarni}, \& {Law}}]{Laher2012APT}
{Laher}, R.~R., {Gorjian}, V., {Rebull}, L.~M., {et~al.} 2012, \pasp, 124, 737,
  \dodoi{10.1086/666883}

\bibitem[{{Lang} {et~al.}(2010){Lang}, {Hogg}, {Mierle}, {Blanton}, \&
  {Roweis}}]{lang2010astrometrynet}
{Lang}, D., {Hogg}, D.~W., {Mierle}, K., {Blanton}, M., \& {Roweis}, S. 2010,
  \aj, 139, 1782, \dodoi{10.1088/0004-6256/139/5/1782}

\bibitem[{{Levison} \& {Duncan}(1994)}]{1994Icar..108...18L}
{Levison}, H.~F., \& {Duncan}, M.~J. 1994, \icarus, 108, 18,
  \dodoi{10.1006/icar.1994.1039}

\bibitem[{{Liu} \& {Ip}(2019)}]{2019ApJ...880...71L}
{Liu}, P.-Y., \& {Ip}, W.-H. 2019, Astrophysical Journal, 880, 71,
  \dodoi{10.3847/1538-4357/ab29eb}

\bibitem[{{Luna} {et~al.}(2014){Luna}, {Satorre}, {Santonja}, \&
  {Domingo}}]{Luna2014AstroIces}
{Luna}, R., {Satorre}, M.~{\'A}., {Santonja}, C., \& {Domingo}, M. 2014, \aap,
  566, A27, \dodoi{10.1051/0004-6361/201323249}

\bibitem[{{Mazzotta Epifani} {et~al.}(2017){Mazzotta Epifani}, {Perna},
  {Dotto}, {Palumbo}, {Dall'Ora}, {Micheli}, {Ieva}, \&
  {Perozzi}}]{2017A&A...597A..59M}
{Mazzotta Epifani}, E., {Perna}, D., {Dotto}, E., {et~al.} 2017, \aap, 597,
  A59, \dodoi{10.1051/0004-6361/201628405}

\bibitem[{{Meech} \& {Belton}(1990)}]{Meech1990ChironAtmosphere}
{Meech}, K.~J., \& {Belton}, M. J.~S. 1990, \aj, 100, 1323,
  \dodoi{10.1086/115600}

\bibitem[{{Mommert}(2017)}]{Mommert2017photometrypipeline}
{Mommert}, M. 2017, Astronomy and Computing, 18, 47,
  \dodoi{10.1016/j.ascom.2016.11.002}

\bibitem[{{Morbidelli}(2008)}]{2008tnoc.book...79M}
{Morbidelli}, A. 2008, {Comets and Their Reservoirs: Current Dynamics and
  Primordial Evolution}, 132

\bibitem[{{Moskovitz} {et~al.}(2019){Moskovitz}, {Schottland}, {Burt},
  {Wasserman}, {Mommert}, {Bailen}, \& {Grimm}}]{AstOrb2019EPSCDPS}
{Moskovitz}, N., {Schottland}, R., {Burt}, B., {et~al.} 2019, in EPSC-DPS Joint
  Meeting 2019, Vol. 2019, EPSC--DPS2019--644

\bibitem[{{Ochsenbein} {et~al.}(2000){Ochsenbein}, {Bauer}, \&
  {Marcout}}]{Ochsenbein2000Vizier}
{Ochsenbein}, F., {Bauer}, P., \& {Marcout}, J. 2000, \aaps, 143, 23,
  \dodoi{10.1051/aas:2000169}

\bibitem[{Oterma(1942)}]{Oterma1942CometOterma}
Oterma, L. 1942, Bureau Central Astronomique de l’Union Astronomique
  Internationale, 900

\bibitem[{{Peixinho} {et~al.}(2020){Peixinho}, {Thirouin}, {Tegler}, {Di
  Sisto}, {Delsanti}, {Guilbert-Lepoutre}, \&
  {Bauer}}]{Peixinho2020CentaursReview}
{Peixinho}, N., {Thirouin}, A., {Tegler}, S.~C., {et~al.} 2020, {From Centaurs
  to Comets - 40 years}, ed. D.~{Prialnik}, M.~A. {Barucci}, \& L.~{Young},
  307--329

\bibitem[{Pravdo {et~al.}(2001)Pravdo, Helin, Hicks, \&
  Lawrence}]{2001IAUC.7738....1P}
Pravdo, S., Helin, E.~F., Hicks, M., \& Lawrence, K. 2001, International
  Astronomical Union Circular, 7738, 1

\bibitem[{Read \& Scotti(2005)}]{Read2005P2005T3IAUC}
Read, M.~T., \& Scotti, J.~V. 2005, International Astronomical Union Circular,
  8614, 2

\bibitem[{{Rein} {et~al.}(2019){Rein}, {Hernandez}, {Tamayo}, {Brown},
  {Eckels}, {Holmes}, {Lau}, {Leblanc}, \& {Silburt}}]{10.1093/mnras/stz769}
{Rein}, H., {Hernandez}, D.~M., {Tamayo}, D., {et~al.} 2019, \mnras, 485, 5490,
  \dodoi{10.1093/mnras/stz769}

\bibitem[{Romanishin {et~al.}(2004)Romanishin, Tegler, Boattini, De~Luise, \&
  Di~Paola}]{Romanishin2004act167P}
Romanishin, W., Tegler, S.~C., Boattini, A., De~Luise, F., \& Di~Paola, A.
  2004, Central Bureau for Astronomical Telegrams, 8545

\bibitem[{{Schorghofer}(2008)}]{WOS:000257834000059}
{Schorghofer}, N. 2008, \apj, 682, 697, \dodoi{10.1086/588633}

\bibitem[{Schwassmann \& Wachmann(1927)}]{SW1927SW1}
Schwassmann, A., \& Wachmann, A.~A. 1927, Bureau Central Astronomique de
  l’Union Astronomique Internationale Observatoire de Copenhague, 171

\bibitem[{{Shi} {et~al.}(2019){Shi}, {Ma}, {Liang}, \&
  {Xu}}]{2019NatSR...9.5492S}
{Shi}, J., {Ma}, Y., {Liang}, H., \& {Xu}, R. 2019, Scientific Reports, 9,
  5492, \dodoi{10.1038/s41598-019-41880-0}

\bibitem[{{Shi} \& {Ma}(2015)}]{2015MNRAS.454.3635S}
{Shi}, J.~C., \& {Ma}, Y.~H. 2015, \mnras, 454, 3635,
  \dodoi{10.1093/mnras/stv2274}

\bibitem[{{Swings}(1986)}]{1986IAUTB..19.....S}
{Swings}, J.~P. 1986, Transactions of the International Astronomical Union,
  Series B, 19

\bibitem[{{Tegler} {et~al.}(2008){Tegler}, {Bauer}, {Romanishin}, \&
  {Peixinho}}]{2008ssbn.book..105T}
{Tegler}, S.~C., {Bauer}, J.~M., {Romanishin}, W., \& {Peixinho}, N. 2008,
  {Colors of Centaurs}, ed. M.~A. {Barucci}, H.~{Boehnhardt}, D.~P.
  {Cruikshank}, A.~{Morbidelli}, \& R.~{Dotson}, 105

\bibitem[{{Tegler} {et~al.}(2016){Tegler}, {Romanishin}, {Consolmagno}, \&
  {J.}}]{Tegler2016colors}
{Tegler}, S.~C., {Romanishin}, W., {Consolmagno}, G.~J., \& {J.}, S. 2016, \aj,
  152, 210, \dodoi{10.3847/0004-6256/152/6/210}

\bibitem[{Tholen {et~al.}(2015)Tholen, Trujillo, \&
  Sheppard}]{Tholen2015P2015T5CBET}
Tholen, D.~J., Trujillo, C., \& Sheppard, S.~S. 2015, Central Bureau Electronic
  Telegrams, 4177

\bibitem[{Wainscoat {et~al.}(2011)Wainscoat, Denneau, Hsieh, Primak, Schultz,
  Watters, Thiel, Goggia, Micheli, Forshay, Lister, Sato, Tholen, Elliott, \&
  Williams}]{Wainscoat2011P2011P2IAUC}
Wainscoat, R., Denneau, L., Hsieh, H., {et~al.} 2011, International
  Astronomical Union Circular, 9225, 1

\bibitem[{Wainscoat {et~al.}(2013)Wainscoat, Veres, Micheli, Denneau, Bolin,
  Tholen, Sato, Guido, Rochowicz, Howes, \&
  Williams}]{Wainscoat2013C2013P4CBET}
Wainscoat, R., Veres, P., Micheli, M., {et~al.} 2013, Central Bureau Electronic
  Telegrams, 3638, 1

\bibitem[{{Willmer}(2018)}]{2018ApJS..236...47W}
{Willmer}, C. N.~A. 2018, \apjs, 236, 47, \dodoi{10.3847/1538-4365/aabfdf}

\bibitem[{{Womack} {et~al.}(2017){Womack}, {Sarid}, \&
  {Wierzchos}}]{womack2017co}
{Womack}, M., {Sarid}, G., \& {Wierzchos}, K. 2017, \pasp, 129, 031001,
  \dodoi{10.1088/1538-3873/129/973/031001}

\end{thebibliography}


\appendix{}

\section{Activity Observation details}
\label{sec:observationdetails}
\begin{center}
	\begin{tabular}{cccrc}
		\# & Instrument  & Date/Time			& Exp.	& Filter\\
		 &       & (UT)				& [s]\ 	&			 \\
		\hline
		 1   & DECam$^1$   & 2017-07-18 09:27	& 137	& $z$	\\
		 2   & DECam$^1$   & 2017-07-18 10:20	& 250	& $z$	\\
		 3   & DECam$^1$   & 2017-07-22 05:37	&  79	& $g$	\\
		 4   & DECam$^1$   & 2017-07-25 06:25	&  60	& $r$	\\
		 5   & DECam$^1$   & 2017-07-25 06:32	&  52	& $r$	\\
		 6   & DECam$^1$   & 2017-08-20 04:48	&  67	& $r$	\\
		 7   & DECam$^2$   & 2019-08-30 09:54	& 250	& \textit{VR}\\
		 8   & DECam$^2$   & 2019-08-30 09:58	& 250	& \textit{VR}\\
		 9   & DECam$^2$   & 2019-08-30 10:03	& 250	& \textit{VR}\\
		10   & DECam$^2$   & 2019-08-30 10:08	& 250	& \textit{VR}\\
		11   & IMACS   & 2019-12-27 00:54  & 300   & WB4800-7800\\
		12   & IMACS   & 2019-12-27 01:01  & 300   & WB4800-7800\\
		13   & IMACS   & 2019-12-27 01:36  & 600   & WB4800-7800\\
		14   & LMI     & 2019-12-30 02:08    & 300   & $g$\\
		15   & LMI     & 2019-12-30 02:17    & 300   & $r$\\
		16   & LMI     & 2019-12-30 02:23    & 300   & $i$\\
		17   & LMI     & 2019-12-30 02:29    & 300   & $g$\\
		18   & LMI     & 2019-12-30 02:35    & 300   & $r$\\
		19   & LMI     & 2019-12-30 02:41    & 300   & $i$\\
	\end{tabular}
	
	$^1$Program 2014B-0404 (PI: Schlegel)\\
	$^2$Program 2019A-0337 (PI: Trilling)\\
\end{center}


\clearpage

\section{Thumbnail Gallery}
\label{sec:thumbnailgallery}

\subsection{Archival Images}
\label{sec:archivalimages}

\begin{center}
	\begin{tabular}{ccc}
	\includegraphics[width=0.32\linewidth]{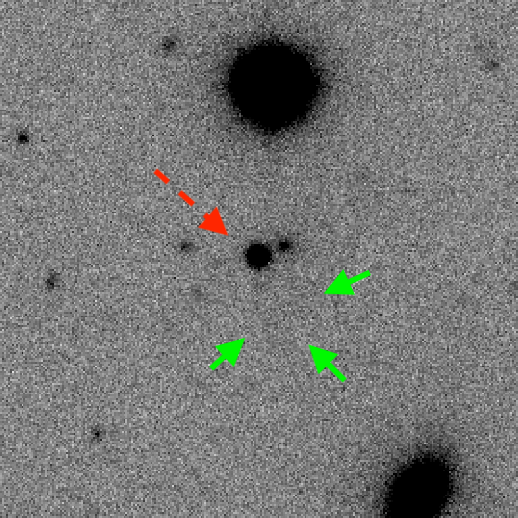} &
	\includegraphics[width=0.32\linewidth]{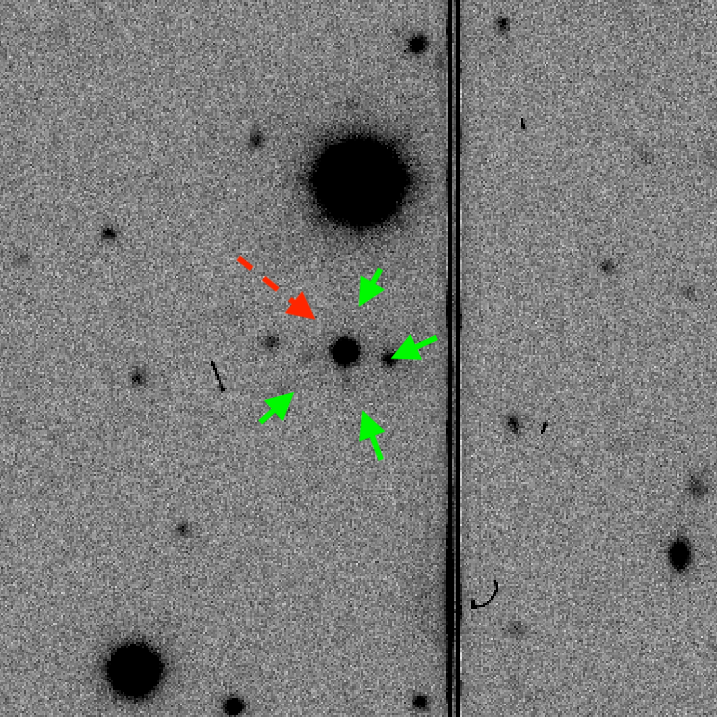} &
	\includegraphics[width=0.32\linewidth]{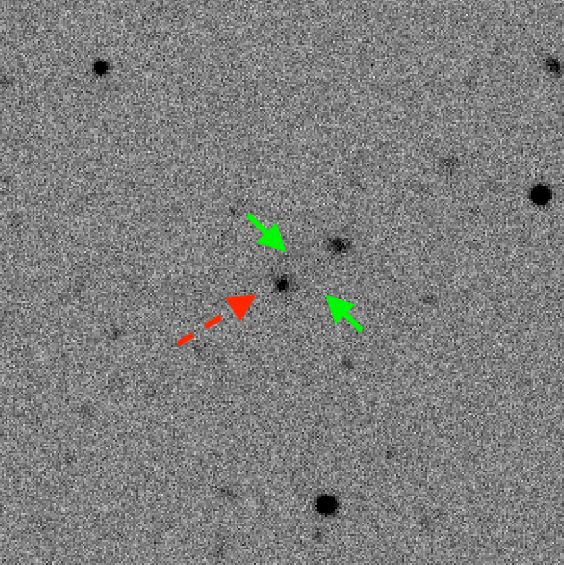}\\
 	\includegraphics[width=0.32\linewidth]{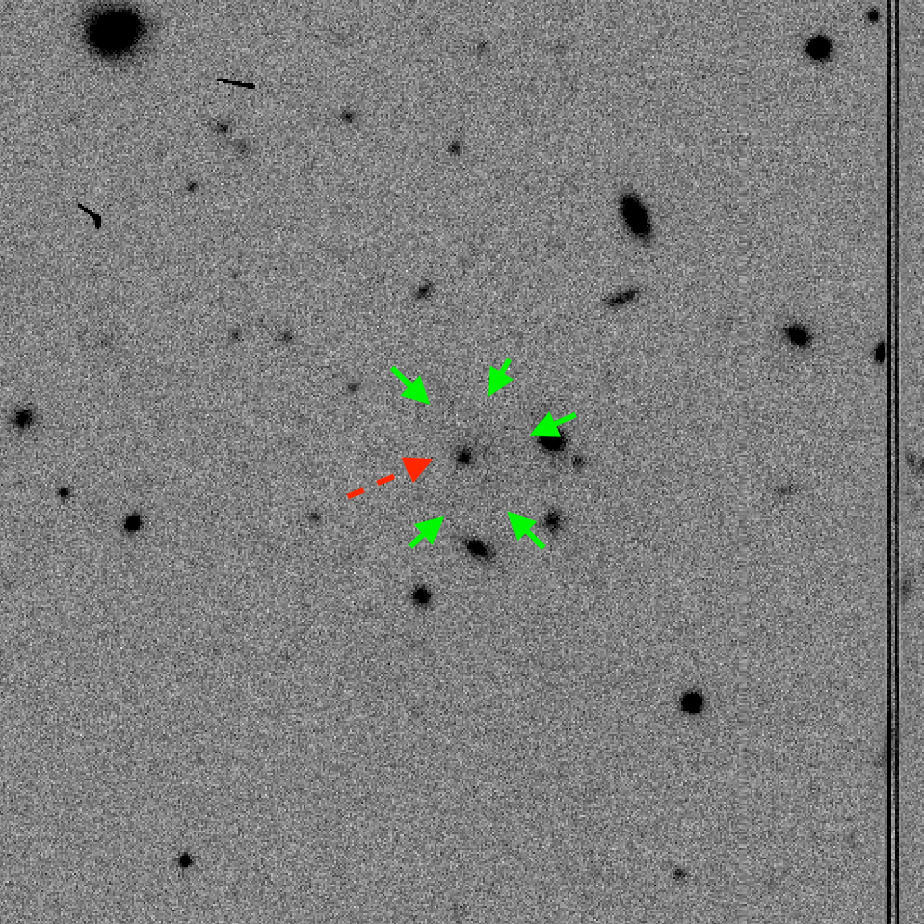} &
 	\includegraphics[width=0.32\linewidth]{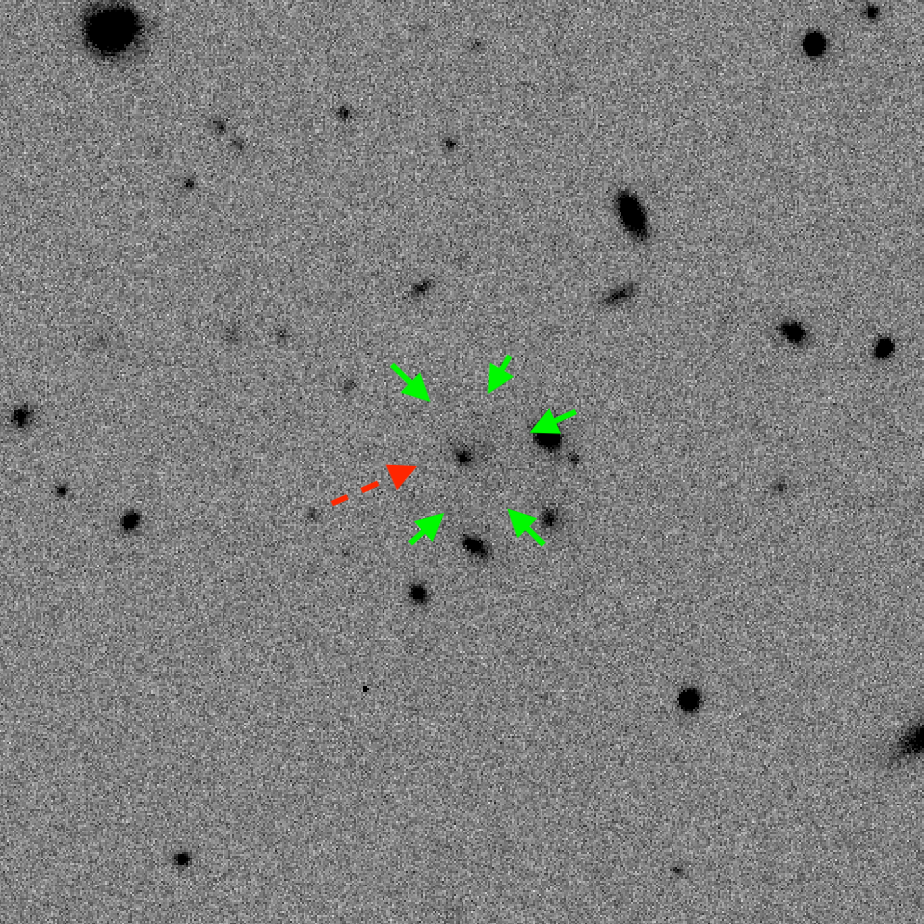} &
 	\includegraphics[width=0.32\linewidth]{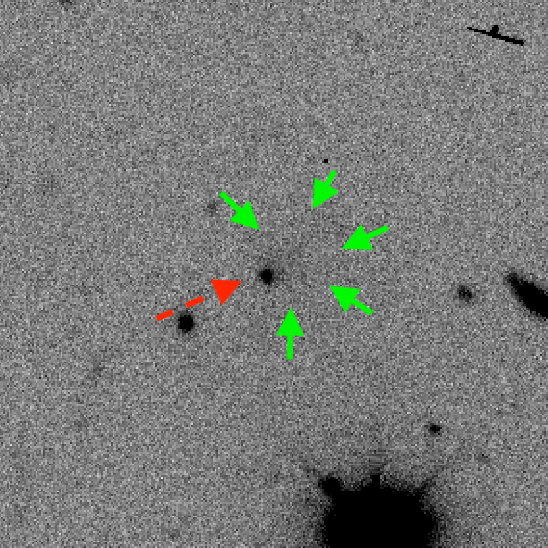} \\
	\end{tabular}
\end{center}

\noindent \textit{DECam Archival Images}: \small{\textbf{Top-left}: UT 2017-Jul-18 09:27 -- 137~s $z$-band \textbf{Top-center}: UT 2017-Jul-18 10:20 -- 250~s $z$-band \textbf{Top-right}: UT 2017-Jul-22 05:37 -- 79~s $g$-band \textbf{Bottom-left}: UT 2017-Jul-25 06:25 -- 60~s $r$-band \textbf{Bottom-center}: UT 2017-Jul-25 06:32 -- 52~s $r$-band \textbf{Bottom-right}: UT 2017-Aug-20 04:48 -- 67~s $r$-band \textbf{All Images}: The coma (green arrows) was exceptionally faint in all of these DECam archival images of \og{} (indicated by dashed red arrows) but nevertheless they prompted us to obtain followup observations.}

\clearpage

\subsection{New DECam Observations}
\label{sec:newobserations}

\begin{center}
	\begin{tabular}{cc}
    \includegraphics[width=0.4\linewidth]{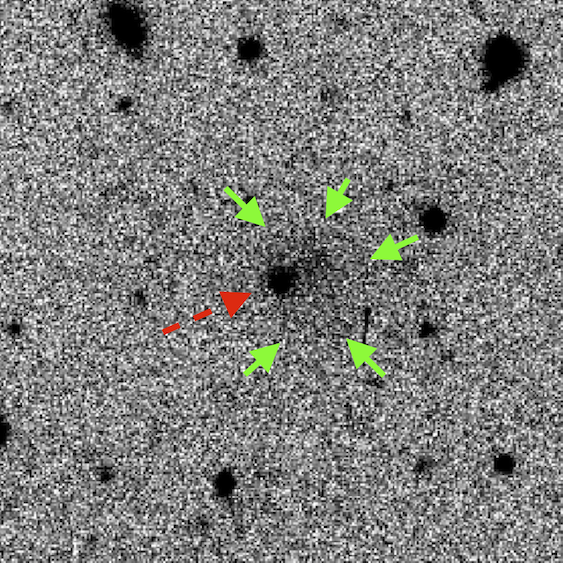} & \includegraphics[width=0.4\linewidth]{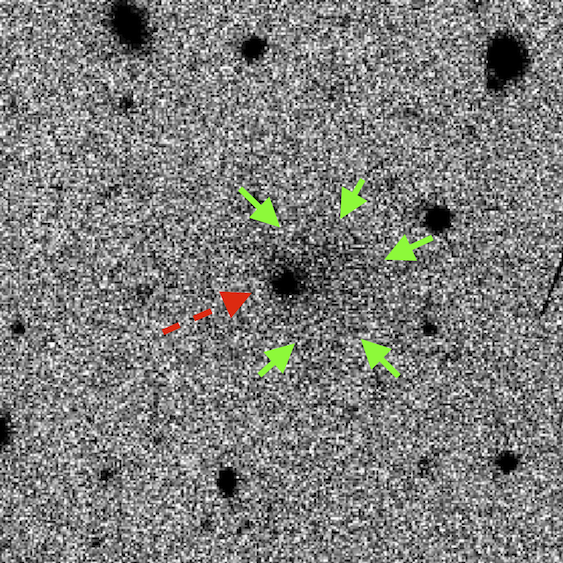}\\
    \includegraphics[width=0.4\linewidth]{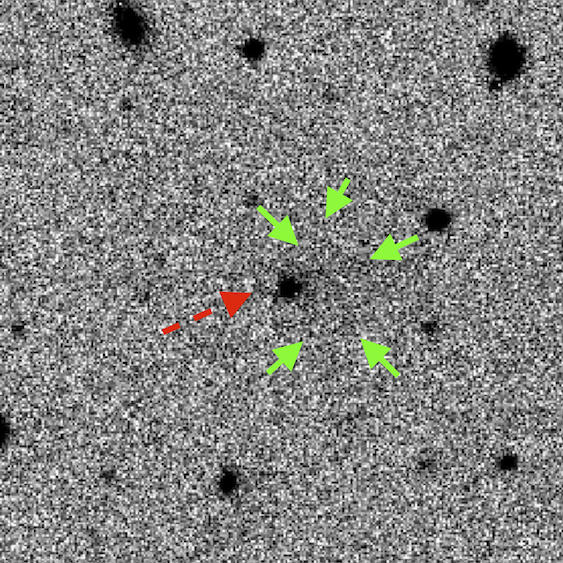} & \includegraphics[width=0.4\linewidth]{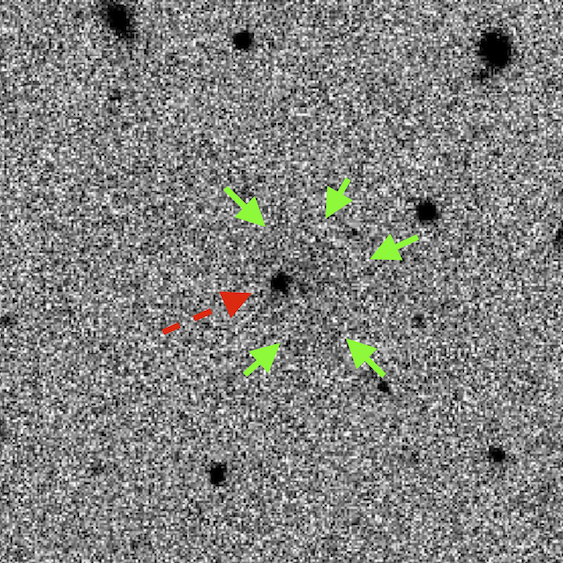}
	\end{tabular}
\end{center}

\noindent \textit{New DECam Observations Gallery}: \textbf{Upper-left}: UT 9:54; \textbf{Upper-right}: UT 9:58; \textbf{Lower-left}: UT 10:03; \textbf{Lower-right}: UT 10:08. \textbf{All images}: (1) dashed red arrow points to \og{}, (2) green arrows highlight the comae if visible, (3) observing date was UT 2019 August 30, (4) filter was \textit{VR}, (5) exposure time was 250 s. The apparent decrease in coma prominence was the result of increasing background noise as images were taken into twilight. 

\clearpage

\subsection{Isophotal Contours}
\label{sec:isophotalcontours}

\begin{center}
	\begin{tabular}{cc}
		\includegraphics[width=0.49\linewidth]{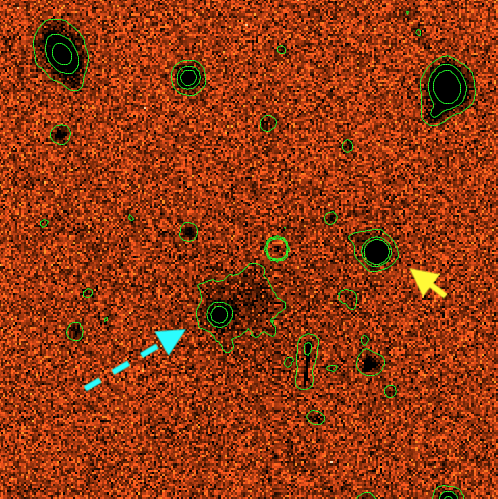} & \includegraphics[width=0.49\linewidth]{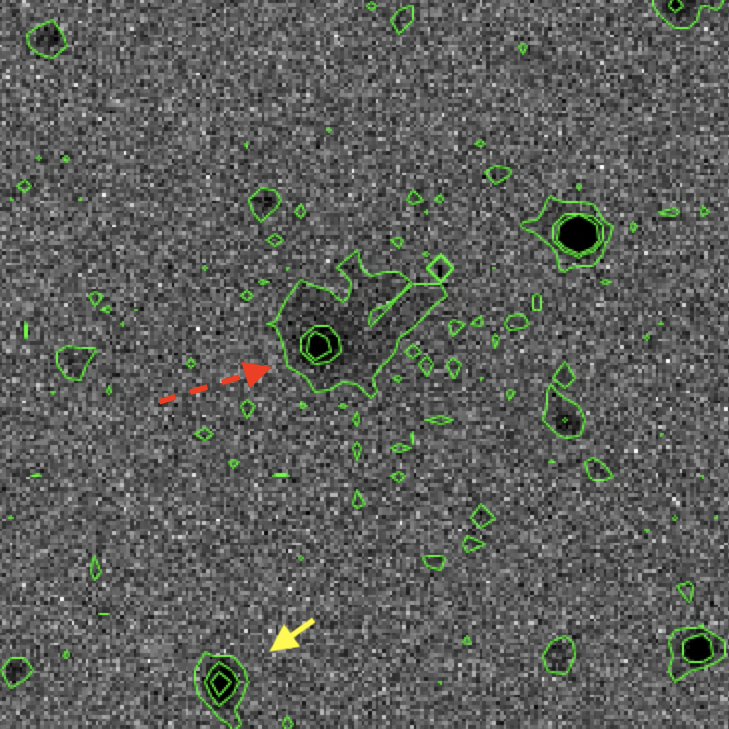}\\
	\end{tabular}
\end{center}

\noindent Isophotal contours indicate the extent and irregularity of the coma emanating from \og{} (dashed arrows), especially when contrasted with background objects (yellow arrows) presenting relatively symmetric radial profiles. These two 250 s \textit{VR}--band exposures were taken at 9:54 (left) and 9:58 (right) during our UT 2019 August 30 followup campaign.

\subsection{New Magellan Observations}
\label{sec:magellanobservations}
\begin{center}
    \begin{tabular}{ccc}
        \includegraphics[width=0.32\linewidth]{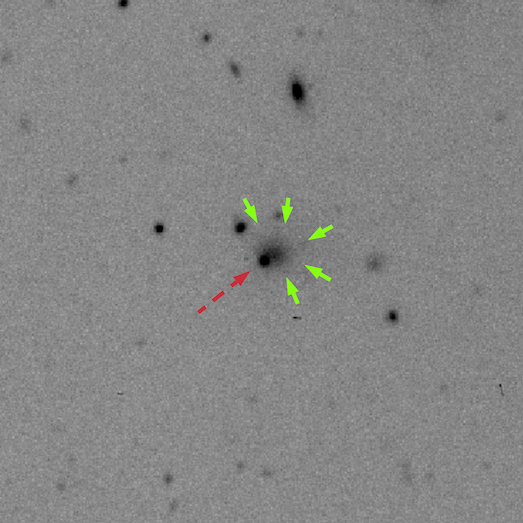} & \includegraphics[width=0.32\linewidth]{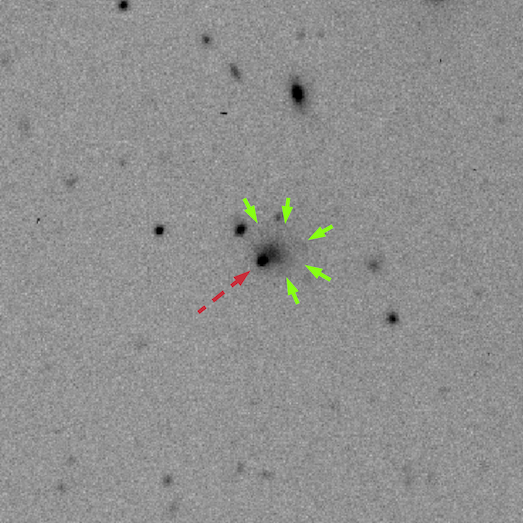} & \includegraphics[width=0.32\linewidth]{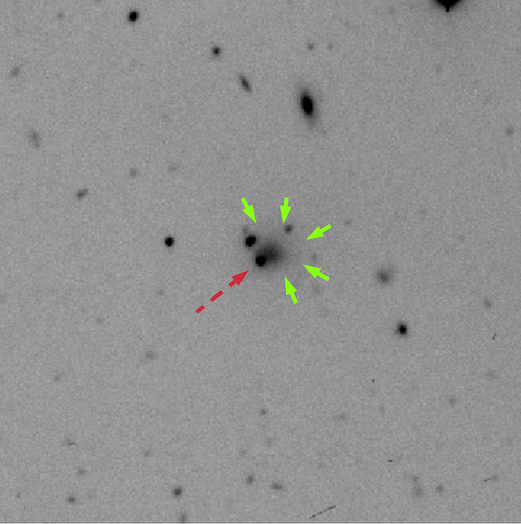} \\
    \end{tabular}
\end{center}

\noindent \og{} imaged December 27, 2019 via the Magellan 6.5m Baade Telescope using the WB4800-7800 filter on the Inamori-Magellan Areal Camera \& Spectrograph (IMACS) at Las Campanas Observatory on Cerro Manqui, Chile. The three images reveal an apparent coma (green arrows) emerging from the object (red dashed arrow) and were taken at 300~s (left, center) exposures and one 600~s exposure (right).

\subsection{New DCT Observations}
\label{sec:dctobservations}

\begin{center}
    \begin{tabular}{ccc}
        \includegraphics[width=0.32\linewidth]{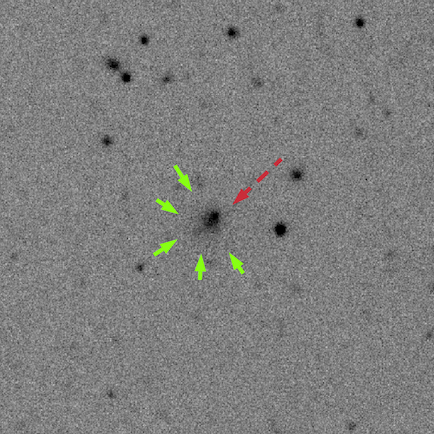} & \includegraphics[width=0.32\linewidth]{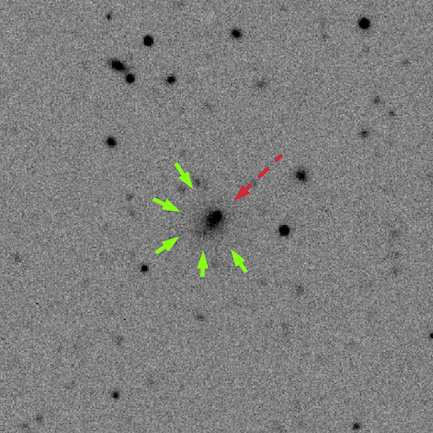} & \includegraphics[width=0.32\linewidth]{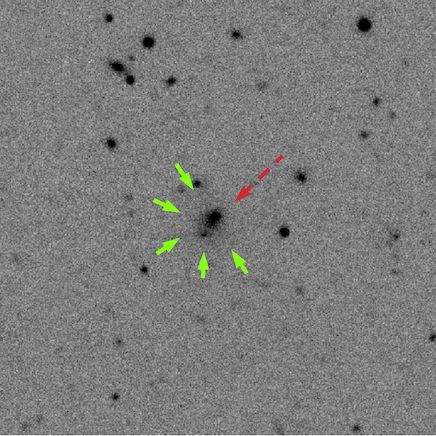} \\
        \includegraphics[width=0.32\linewidth]{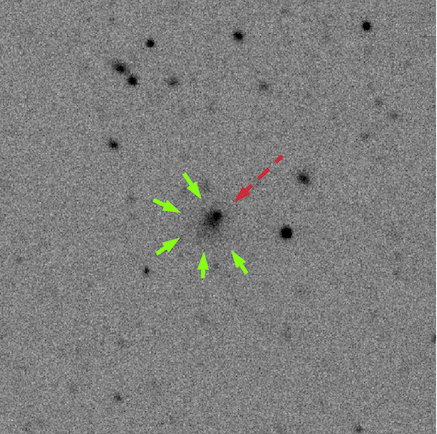} & \includegraphics[width=0.32\linewidth]{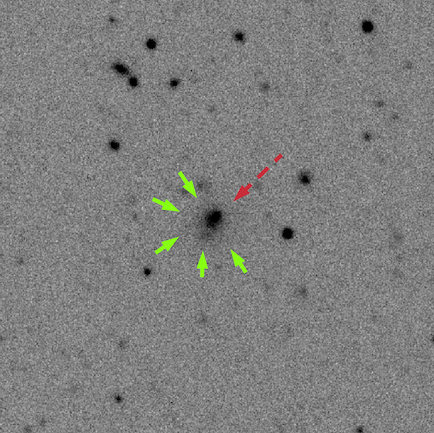} & \includegraphics[width=0.32\linewidth]{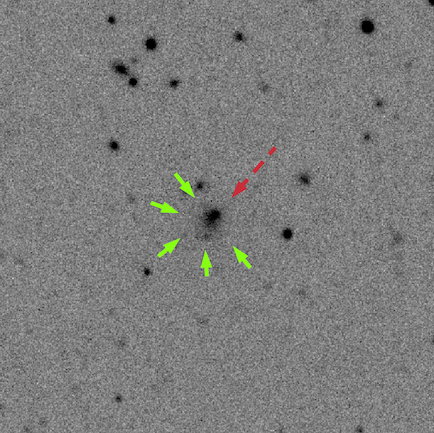} \\
    \end{tabular}
\end{center}

\noindent \og{} imaged December 30, 2019, via the Lowell Observatory 4.3~m Discovery Channel Telescope (Arizona, USA) using the Large Monolithic Imager (LMI). Green arrows trace out a diffuse coma and a dashed red arrow points to the nucleus in each of the six images. Each exposure in the two $g$-$r$-$i$ sequences (top and bottom rows) was 300~s long.


\end{document}